\documentclass[conference]{IEEEtran}
\usepackage{amssymb}
\usepackage[cmex10]{amsmath}
\usepackage{stfloats}
\usepackage{graphicx}
\usepackage{subfigure}
\usepackage{tabularx}
\usepackage{epsfig,epsf,color,balance,cite}
\usepackage{verbatim}
\usepackage{url}
\usepackage{bm}
\usepackage{mathrsfs}
\usepackage{amsthm}
\usepackage{mathtools}
\usepackage{extarrows}
\DeclarePairedDelimiter{\ceil}{\lceil}{\rceil}
\newtheorem{theorem}{Theorem}

\newtheorem{lemma}{Lemma}

\usepackage{algorithm}
\usepackage{algpseudocode}
\usepackage{graphicx}
\usepackage{epstopdf}
\begin{document}

\title{Information Bottleneck for an Oblivious Relay with Channel State Information: the Vector Case}

\author{
	\IEEEauthorblockN{Hao Xu\IEEEauthorrefmark{1},
		Tianyu Yang\IEEEauthorrefmark{1},
		Giuseppe Caire\IEEEauthorrefmark{1},
		and
		Shlomo Shamai (Shitz)\IEEEauthorrefmark{2}
	}
\IEEEauthorblockA{\IEEEauthorrefmark{1}Faculty of Electrical Engineering and Computer Science, Technical University of Berlin, 10587 Berlin, Germany}
\IEEEauthorblockA{\IEEEauthorrefmark{2}Department of Electrical Engineering, Technion—Israel Institute of Technology, Haifa 3200003, Israel}
\IEEEauthorblockA{E-mail: xuhao@mail.tu-berlin.de; tianyu.yang@tu-berlin.de; caire@tu-berlin.de; sshlomo@ee.technion.ac.il}
}

\maketitle

\begin{abstract}
This paper considers the information bottleneck (IB) problem of a Rayleigh fading multiple-input multiple-out (MIMO) channel.
Due to the bottleneck constraint, it is impossible for the oblivious relay to inform the destination node of the perfect channel state information (CSI) in each channel realization.
To evaluate the bottleneck rate, we provide an upper bound by assuming that the destination node can get the perfect CSI at no cost and two achievable schemes with simple symbol-by-symbol relay processing and compression.
Numerical results show that the lower bounds obtained by the proposed achievable schemes can come close to the upper bound on a wide range of relevant system parameters.
\end{abstract}

%\begin{keywords}
%	information bottleneck (IB), oblivious relay, Rayleigh fading, source coding, quantization.
%\end{keywords}

\IEEEpeerreviewmaketitle

%%%%%%%%%%%%%%%%%%%%%%%%%%%%%%%%%%%%%%%%%%%%%%%%
\section{Introduction}
\label{introduction}

For a Markov chain $X \rightarrow Y \rightarrow Z$ and an assigned joint probability distribution $p_{X,Y}$, consider the following information bottleneck (IB) problem 
\begin{subequations}\label{IB_problem0}
	\begin{align}
	\mathop {\max }\limits_{p_{Z|Y}} \quad & I(X; Z) \label{IB_problem0_a}\\
	\text{s.t.} \quad\;\; &  I(Y; Z) \leq C, \label{IB_problem0_b}
	\end{align}
\end{subequations}
where $C$ is the bottleneck constraint parameter and the optimization is with respect to the conditional probability distribution $p_{Z|Y}$ of $Z$ given $Y$.
Formulation (\ref{IB_problem0}) was introduced by Tishby in \cite{tishby2000information}, and has been used to interpret the behavior of deep learning neural networks \cite{shwartz2017opening}.
From a more fundamental information theoretic viewpoint, the IB arises from the classical remote source coding problem \cite{dobrushin1962information, witsenhausen1980indirect} under logarithmic distortion \cite{courtade2013multiterminal}.

An interesting application of the IB problem in communications consists of a source node, an oblivious relay, and a destination node, which is connected to the relay via an error-free link with capacity $C$.
The source node sends codewords over a communication channel and an observation is made at the relay.
$X$ and $Y$ are respectively the channel input from the source node and output at the relay.
The relay is oblivious in the sense that it cannot decode the information message of the source node itself.
This feature can be modeled rigorously by assuming that the source and destination nodes make use of a codebook selected at random over a library, while the relay is unaware of such random selection. 
For example, in a cloud radio access network (C-RAN), each remote radio head (RRH) acts as a relay and is usually constrained to implement only radio functionalities while the baseband functionalities are migrated to the cloud central processor, particularly as the network size gets large \cite{aguerri2019capacity}.

Due to the oblivious feature, the relaying strategies which require the codebooks to be known at the relay, e.g., decode-and-forward, compute-and-forward, etc. \cite{nazer2011compute, hong2013compute, nazer2009structured} cannot be applied.
Instead, the relay has to perform oblivious processing, i.e., employ strategies in forms of compress-and-forward \cite{simeone2011codebook, park2012robust, zhou2016optimal, aguerri2016lossy}.
In particular, the relay must treat $X$ as a random process, produce some useful representation $Z$, and convey it to the destination node subject to the link constraint $C$. 
Then, it makes sense to find $Z$ such that $I(X; Z)$ is maximized.

The IB problem for this kind of communication scenario has been studied in \cite{demel2020cloud, winkelbauer2014rate, winkelbauer2014ratevec, caire2018information}.
In \cite{demel2020cloud}, the IB method was applied to reduce the fronthaul data rate of a C-RAN network.
References \cite{winkelbauer2014rate} and \cite{winkelbauer2014ratevec} respectively considered Gaussian scalar and vector channels with IB constraint, and investigated the optimal trade-off between the compression rate and the relevant information.
However, all references \cite{demel2020cloud}, \cite{winkelbauer2014rate}, and \cite{winkelbauer2014ratevec} considered block fading channels, and assumed that the perfect channel state information (CSI) was known at both the relay and the destination node.
In \cite{caire2018information}, the IB problem of a scalar Rayleigh fading channel was studied.
Due to the bottleneck constraint, it is impossible to inform the destination node of the perfect CSI in each channel realization.
An upper bound and two achievable schemes were provided in \cite{caire2018information}.

In this paper, we extend the work in \cite{caire2018information} to the multiple-input multiple-out (MIMO) channel  with independent and identically distributed (i.i.d.) Rayleigh fading.
To evaluate the bottleneck rate, we first obtain an upper bound by assuming that the channel matrix is also known at the destination node with no cost.
Then, we provide two achievable schemes where the first scheme transmits the compressed noisy signal as well as the quantized noise levels to the destination node, while the second scheme only transmits a compressed estimate.
Numerical results show that with simple symbol-by-symbol relay processing and compression, the lower bounds obtained by the proposed achievable schemes can come close to the upper bound on a wide range of relevant system parameters.

%%%%%%%%%%%%%%%%%%%%%%%%%%%%%%%%%%%%%%%%%%%%%%%%%%
\section{Problem Formulation}
\label{problem_formu}

We consider a system with a source node, an oblivious relay, and a destination node.
For convenience, we call the source-relay channel, `Channel~1', and the relay-destination channel, `Channel~2'. 
For Channel~1, we consider the following Gaussian MIMO channel with i.i.d. Rayleigh fading 
\begin{equation}\label{relay_obser}
\bm y = \bm H \bm x + \bm n,
\end{equation}
where $\bm x \in {\mathbb C}^{K \times 1}$ and $\bm n \in {\mathbb C}^{M\times1}$ are respectively zero-mean circularly symmetric complex Gaussian input and noise with covariance matrices $\bm I_K$ and $\sigma^2 \bm I_M$, i.e., ${\bm x} \sim {\cal CN}(\bm 0, \bm I_K)$ and ${\bm n} \sim {\cal CN}(\bm 0, \sigma^2 \bm I_M)$.
$\bm H \in {\mathbb C}^{M \times K}$ is a random matrix independent of both $\bm x$ and $\bm n$, and the elements of $\bm H$ are i.i.d. zero-mean unit-variance complex Gaussian random variables, i.e., $\bm H \sim {\cal CN} (\bm 0, \bm I_K \otimes \bm I_M)$.
Let $\rho = \frac{1}{\sigma^2}$ denote the signal-to-noise ratio (SNR).
Let $\bm z$ denote a useful representation of $\bm y$ produced by the relay for the destination node.
$\bm x \rightarrow (\bm y, \bm H) \rightarrow \bm z$ thus forms a Markov chain.
We assume that the relay node has a direct observation of the channel matrix $\bm H$, while the destination node does not.
Then, we consider the following IB problem
\begin{subequations}\label{IB_problem}
	\begin{align}
	\mathop {\max }\limits_{p(\bm z| \bm y, \bm H)} \quad & I(\bm x; \bm z) \label{IB_problem_a}\\
	\text{s.t.} \quad\;\;\; &  I(\bm y, \bm H; \bm z) \leq C, \label{IB_problem_b}
	\end{align}
\end{subequations}
where $C$ is the bottleneck constraint, i.e., the link capacity of Channel~2.
In this paper, we call $I(\bm x; \bm z)$ the bottleneck rate and $I(\bm y, \bm H; \bm z)$ the compression rate.
Obviously, for a joint probability distribution $p(\bm x, \bm y, \bm H)$ determined by (\ref{relay_obser}), problem (\ref{IB_problem}) is a slightly augmented version of IB problem (\ref{IB_problem0}).
In our problem, we aim to find a conditional distribution $p(\bm z| \bm y, \bm H)$ such that bottleneck constraint (\ref{IB_problem_b}) is satisfied and the bottleneck rate is maximized, i.e., as much as information of $\bm x$ can be extracted from representation $\bm z$.

\section{Informed Receiver Upper Bound}
\label{informed_ub}

As stated in \cite{caire2018information}, an obvious upper bound to problem (\ref{IB_problem}) can be obtained by letting both the relay and the destination node know the channel matrix $\bm H$.
We call the bound in this case the informed receiver upper bound.
The IB problem in this case takes on the following form
\begin{subequations}\label{IB_problem1}
	\begin{align}
	\mathop {\max }\limits_{p(\bm z| \bm y, \bm H)} \quad & I(\bm x; \bm z| \bm H) \label{IB_problem1_a}\\
	\text{s.t.} \quad\;\; &  I(\bm y; \bm z| \bm H) \leq C. \label{IB_problem1_b}
	\end{align}
\end{subequations}
In \cite{winkelbauer2014rate}, the IB problem for a scalar Gaussian channel with block fading has been studied.
In the following theorem, we show that for the considered MIMO channel with Rayleigh fading, (\ref{IB_problem1}) can be decomposed into a set of parallel scalar IB problems, and the informed receiver upper bound can be obtained based on the result in \cite{winkelbauer2014rate}.

\begin{theorem}\label{theorem_ub}
For the considered MIMO channel with Rayleigh fading, the informed receiver upper bound is
\begin{equation}\label{R_up_KM}
R^{\text {ub}} = T \int_{\frac{\nu}{\rho}}^{\infty} \left[ \log \left(1 + \rho \lambda\right) - \log (1 + \nu)\right] f_\lambda (\lambda) d \lambda,
\end{equation}
where $T = \min \{K, M\}$, the probability density function (pdf) of $\lambda$, i.e., $f_\lambda (\lambda)$, is given by (\ref{pdf}), and $\nu$ is chosen such that the following bottleneck constraint is met
\begin{equation}\label{bottle_constr_KM}
\int_{\frac{\nu}{\rho}}^{\infty} \left( \log \frac{\rho \lambda}{ \nu} \right) f_\lambda (\lambda) d \lambda = \frac{C}{T}.
\end{equation}
\end{theorem}

\itshape \textbf{Proof:}  \upshape
	See Appendix \ref{prove_theorem_ub}.
\hfill $\Box$

\begin{lemma}\label{lemma_ub}
	When $M \rightarrow + \infty$ or $\rho \rightarrow + \infty$, upper bound $R^{\text {ub}}$ tends asymptotically to $C$.
	When $C \rightarrow + \infty$, $R^{\text {ub}}$ approaches the capacity of Channel~1, i.e.,
	\begin{align}\label{R_up_appro}
	R^{\text {ub}} & \rightarrow I(\bm x; \bm y, \bm H)\nonumber\\
	 & = T \int_{0}^{\infty} \log \left(1 + \rho \lambda\right) f_\lambda (\lambda) d \lambda.
	\end{align}
\end{lemma}

\itshape \textbf{Proof:}  \upshape
	See Appendix \ref{prove_lemma_ub}.
\hfill $\Box$

%%%%%%%%%%%%%%%%%%%%%%%%%%%%%%%%%%%%%%%%%%%%%%%%%%%%%%%%
\section{Achievable Schemes}
\label{achiev_schemes}

In this section, we provide two achievable schemes where each scheme satisfies the bottleneck constraint and gives a lower bound to the bottleneck rate.

\subsection{Quantized channel inversion (QCI) scheme when $K \leq M$}
\label{QCI_scheme}

In our first achievable scheme, the relay first gets an estimate of the channel input using channel inversion and then transmits the quantized noise levels as well as the compressed noisy signal to the destination node.

In particular, we apply the pseudo inverse matrix of $\bm H$, i.e., $(\bm H^H \bm H)^{-1} \bm H^H$, to $\bm y$, and get the zero-forcing estimate of $\bm x$ as follows
\begin{align}\label{ZF_estimate2}
	{\tilde {\bm x}} & = (\bm H^H \bm H)^{-1} \bm H^H \bm y \nonumber\\
	& = \bm x + (\bm H^H \bm H)^{-1} \bm H^H \bm n \nonumber\\
	& \triangleq \bm x + {\tilde {\bm n}}.
\end{align}
For a given channel matrix $\bm H$, ${\tilde {\bm n}} \sim {\cal CN}(\bm 0, \bm A)$, where $\bm A = \sigma^2 (\bm H^H \bm H)^{-1}$.
Let $\bm A = \bm A_1 + \bm A_2$, where $\bm A_1$ and $\bm A_2$ respectively consist of the diagonal and off-diagonal elements of $\bm A$, i.e., $\bm A_1 = \bm A \odot \bm I_K$ and $\bm A_2 = \bm A - \bm A_1$.
If $\bm H$ can be perfectly transmitted to the destination node, the bottleneck rate could be obtained by following similar steps in Appendix~\ref{prove_theorem_ub}.
However, since $\bm H$ follows a non-degenerate continuous distribution and the bottleneck constraint is finite, this is not possible. 
To reduce the number of bits per channel use required for informing the destination node of the channel information, we only convey a compressed version of $\bm A_1$ and consider a set of independent scalar Gaussian sub-channels.

Specifically, we force each diagonal entry of $\bm A_1$ to belong to a finite set of quantized levels by adding artificial noise, i.e., by introducing physical degradation.
We fix a finite grid of $J$ positive quantization points ${\cal B} = \{ b_1, \cdots, b_J \}$, where $b_1 \leq b_2 \leq \cdots \leq b_{J-1} < b_J$, $b_J = + \infty$, and define the following ceiling operation
\begin{equation}\label{ceiling}
\ceil[\big]{a}_{\cal B} = \arg \min_{b \in {\cal B}} \{ a \leq b  \}.
\end{equation}
Then, by adding a Gaussian noise vector ${\tilde {\bm n}}' \sim {\cal CN} \left( \bm 0, \right.$ $\left. {\text {diag}} \left\{ \ceil[\big]{a_1}_{\cal B} - a_1, \cdots, \ceil[\big]{a_K}_{\cal B} - a_K\right\} \right)$, which is independent of everything else, to (\ref{ZF_estimate2}), a degraded version of ${\tilde {\bm x}}$ can be obtained as follows
\begin{align}\label{degraded_Gaussian}
{\hat {\bm x}} & = {\tilde {\bm x}} + {\tilde {\bm n}}'\nonumber\\
& = \bm x + {\tilde {\bm n}} + {\tilde {\bm n}}'\nonumber\\
& \triangleq \bm x + {\hat {\bm n}},
\end{align}
where ${\hat {\bm n}} \sim {\cal CN} \left( \bm 0,  \bm A_1' + \bm A_2\right)$ for a given $\bm H$ and $\bm A_1' \triangleq {\text {diag}} \left\{ \ceil[\big]{a_1}_{\cal B}, \cdots, \ceil[\big]{a_K}_{\cal B}\right\}$.
Obviously, due to $\bm A_2$, the elements in noise vector ${\hat {\bm n}}$ are correlated.

To evaluate the bottleneck rate, we consider a new variable
\begin{equation}\label{x_bar}
{\hat {\bm x}}_g = \bm x + {\hat {\bm n}}_g,
\end{equation}
where ${\hat {\bm n}}_g \sim {\cal CN} \left( \bm 0,  \bm A_1'\right)$.
Obviously, (\ref{x_bar}) can be seen as $K$ parallel scalar Gaussian sub-channels with noise power $\ceil[\big]{a_k}_{\cal B}$ for each sub-channel.
Since each quantized noise level $\ceil[\big]{a_k}_{\cal B}$ only has $J$ possible values, it is possible for the relay to inform the destination node of the channel information via the constrained link.
Note that from the definition of $\bm A$ in (\ref{ZF_estimate2}), it is known that $a_k, ~\forall~ k \in {\cal K} \triangleq \{ 1, \cdots, K \}$ are correlated.
The quantized noise levels $\ceil[\big]{a_k}_{\cal B}, ~\forall~ k \in {\cal K}$ are thus also correlated.
Hence, we can jointly source-encode $\ceil[\big]{a_k}_{\cal B}, ~\forall~ k \in {\cal K}$ to further reduce the number of bits used for CSI feedback.
However, since the joint entropy of the quantization indices is difficult to obtain (even numerically, since it is a discrete joint distribution over $J^K$ possible values), in this work we consider the (slightly) suboptimal, but far more practical, entropy coding of each sub-channel quantization index separately. 
The resulting optimization problem becomes
\begin{subequations}\label{problem_QCI}
	\begin{align}
	\mathop {\max }\limits_{p({\hat {\bm z}}_g| {\hat {\bm x}}_g)} \quad & I(\bm x; {\hat {\bm z}}_g| \bm A_1') \label{problem_QCI_a}\\
	\text{s.t.} \quad\;\;\; &  I({\hat {\bm x}}_g; {\hat {\bm z}}_g| \bm A_1') \leq C - \sum_{k=1}^K H_k, \label{problem_QCI_b}
	\end{align}
\end{subequations}
where $H_k$ denotes the entropy of $\ceil[\big]{a_k}_{\cal B}$.
In Appendix \ref{prove_theorem_QCI}, we show that $a_k, \forall k \in {\cal K}$ are marginally identically inverse chi squared distributed with $M-K+1$ degrees of freedom.
Hence, $H_k = H_0 \triangleq - \sum_{j=1}^J P_j \log P_j$, where $P_j = {\text {Pr}}\left\{ \ceil[\big]{a}_{\cal B} = b_j \right\}$ and $a$ follows the same distribution as $a_k$.
The pdf of $a$ is given in (\ref{pdf_a}), based on which the probability mass function (pmf) $P_j$ can be calculated as (\ref{P_j}).
In the following theorem, we give a lower bound to the bottleneck rate by solving IB problem (\ref{problem_QCI}).

\begin{theorem}\label{theorem_QCI}
	If $\bm A_1'$ is conveyed to the destination node for each channel realization, by solving IB problem (\ref{problem_QCI}), the following lower bound to the bottleneck rate can be obtained
	\begin{equation}\label{R_lb1}
	R^{\text {lb}1} = \sum_{j=1}^{J-1} K P_j \left[ \log \left(1 + \rho_j\right) - \log (1 + \rho_j 2^{-c_j} ) \right].
	\end{equation}
	where $\rho_j = \frac{1}{b_j}$, $c_j = \left[ \log \frac{\rho_j}{\nu} \right]^+$, and $\nu$ is chosen such that the following bottleneck constraint is met
	\begin{equation}\label{BT_constraint}
	\sum_{j=1}^{J-1} K P_j c_j = C - K H_0.
	\end{equation}
\end{theorem}

\itshape \textbf{Proof:}  \upshape
	See Appendix \ref{prove_theorem_QCI}.
\hfill $\Box$

Since (\ref{x_bar}) can be seen as $K$ parallel scalar Gaussian sub-channels, according to \cite[(16)]{winkelbauer2014rate}, the representation of ${\hat {\bm x}}_g$, i.e., ${\hat {\bm z}}_g$, can be constructed by adding independent fading and Gaussian noise to each element of ${\hat {\bm x}}_g$.
Denote
\begin{align}\label{z_bar}
{\hat {\bm z}}_g & = \bm \varPhi {\hat {\bm x}}_g + {\hat {\bm n}}_g' \nonumber\\
& = \bm \varPhi \bm x + \bm \varPhi {\hat {\bm n}}_g + {\hat {\bm n}}_g',
\end{align}
where $\bm \varPhi$ is a diagonal matrix with positive and real diagonal entries, and ${\hat {\bm n}}_g' \sim {\cal CN} \left( \bm 0,  \bm I_K \right)$.
Note that ${\hat {\bm x}}_g$ in (\ref{x_bar}) and its representation ${\hat {\bm z}}_g$ in (\ref{z_bar}) are only auxiliary variables.
What we are really interested in is the representation of ${\hat {\bm x}}$ and the corresponding bottleneck rate.
Hence, we also add fading $\bm \varPhi$ and Gaussian noise ${\hat {\bm n}}_g'$ to ${\hat {\bm x}}$ in (\ref{degraded_Gaussian}) and get its representation as follows
\begin{align}\label{z_QCI}
\bm z & = \bm \varPhi {\hat {\bm x}} + {\hat {\bm n}}_g' \nonumber\\
& = \bm \varPhi \bm x + \bm \varPhi {\hat {\bm n}} + {\hat {\bm n}}_g'.
\end{align}
In the following lemma we show that by transmitting quantized noise levels $\ceil[\big]{a_k}_{\cal B},~\forall k \in {\cal K}$ and representation $\bm z$ to the destination node, $R^{\text {lb}1}$ is an achievable lower bound to the bottleneck rate and the bottleneck constraint is satisfied.
\begin{lemma}\label{lemma_ineq_QCI}
	If $\bm A_1'$ is forwarded to the destination node for each channel realization, with signal vectors ${\hat {\bm x}}$ and ${\hat {\bm x}}_g$ in (\ref{degraded_Gaussian}) and (\ref{x_bar}), and their representations $\bm z$ and ${\hat {\bm z}}_g$ in (\ref{z_QCI}) and (\ref{z_bar}), we have
	\begin{align}
	I({\hat {\bm x}}; \bm z| \bm A_1') & \leq I({\hat {\bm x}}_g; {\hat {\bm z}}_g| \bm A_1'), \label{ineq3}\\
	I(\bm x; \bm z| \bm A_1') & \geq I(\bm x; {\hat {\bm z}}_g| \bm A_1'), \label{ineq4}
	\end{align}
	where (\ref{ineq3}) indicates that $I({\hat {\bm x}}; \bm z| \bm A_1') \leq C - K H_0$ and (\ref{ineq4}) gives $I(\bm x; \bm z| \bm A_1') \geq R^{\text {lb}1}$.
\end{lemma}

\itshape \textbf{Proof:}  \upshape
	See Appendix \ref{prove_lemma_ineq_QCI}.
\hfill $\Box$

\begin{lemma}\label{lemma_QCI}
When $M \rightarrow + \infty$ or $\rho \rightarrow + \infty$, we can always find a sequence of quantization points ${\cal B} = \{ b_1, \cdots, b_J \}$ such that $R^{\text {lb}1} \rightarrow C$.
When $C \rightarrow + \infty$,
\begin{align}
R^{\text {lb}1} & \rightarrow K \mathbb{E} \left[ \log \left( 1 + \frac{1}{a} \right) \right] \nonumber\\
& \leq I(\bm x; \bm y, \bm H),
\end{align}
where the expectation can be calculated by using the pdf of $a$ in (\ref{pdf_a}) and $I(\bm x; \bm y, \bm H)$ is the capacity of Channel~1.
\end{lemma}

\itshape \textbf{Proof:}  \upshape
	See Appendix \ref{prove_lemma_QCI}.
\hfill $\Box$

For the sake of simplicity, we may choose the quantization levels as quantiles such that we obtain the uniform pmf $P_j = \frac{1}{J}$.
The lower bound (\ref{R_lb1}) can thus be simplified as
\begin{equation}
R^{\text {lb}1} = \sum_{j=1}^{J-1} \frac{K}{J} \left[ \log \left(1 + \rho_j\right) - \log (1 + \rho_j 2^{-c_j}) \right],
\end{equation}
and the bottleneck constraint (\ref{BT_constraint}) becomes
\begin{equation}\label{cons}
\sum_{j=1}^{J-1} \left[ \log \frac{\rho_j}{\nu} \right]^+ = \frac{JC}{K} - J B,
\end{equation}
where $B = \log J$ can be seen as the number of bits required for quantizing each diagonal entry of $\bm A_1$.
Since $\rho_1 \geq \cdots \geq \rho_{J-1}$, from the strict convexity of the problem, we know that there must exist a unique integer $1 \leq l \leq J-1$ such that 
\begin{align}\label{nu}
& \sum_{j=1}^l \log \frac{\rho_j}{\nu} = \frac{JC}{K} - J B, \nonumber\\
& \rho_j \leq \nu, ~\forall~ l+1 \leq j \leq J-1.
\end{align}
Hence, $\nu$ can be obtained from
\begin{equation}
\log \nu = \sum_{j=1}^l \frac{\log \rho_j}{l} - \frac{JC}{lK} + \frac{J B}{l},
\end{equation}
and $R^{\text {lb}1}$ can be calculated as follows
\begin{equation}
R^{\text {lb}1} = \sum_{j=1}^l \frac{K}{J} \left[ \log \left(1 + \rho_j\right) - \log (1 + \nu) \right].
\end{equation}
Then, we only need to test the above condition for $l=1, 2, 3, \cdots$ till (\ref{nu}) is satisfied.
Note that to ensure $R^{\text {lb}1} > 0$, $\frac{JC}{K} - J B$ in (\ref{cons}) has to be positive, i.e., $B < \frac{C}{K}$.
Moreover, though choosing the quantization levels as quantiles makes it easier to calculate $R^{\text {lb}1}$, the results in Lemma~\ref{lemma_QCI} may not hold in this case since the choice of quantization points ${\cal B} = \{ b_1, \cdots, b_J \}$ is restricted.

\subsection{MMSE estimate at the relay}
\label{MMSE_scheme}

In the second achievable scheme, we assume that the relay first produces the MMSE estimate of $\bm x$ given $(\bm y, \bm H)$, and then source-encode this estimate.

Denote 
\begin{equation}\label{F}
\bm F = \left( \bm H \bm H^H + \sigma^2 \bm I_M \right)^{-1} \bm H.
\end{equation}
The MMSE estimate of $\bm x$ is thus given by
\begin{align}\label{bar_y_lb4}
{\bar {\bm x}} & =  \bm F^H \bm y\nonumber\\
& = \bm F^H \bm H \bm x + \bm F^H \bm n.
\end{align}
Then, we consider the following modified IB problem
\begin{subequations}\label{IB_problem_lb1}
	\begin{align}
	\mathop {\max }\limits_{p(\bm z| {\bar {\bm x}})} \quad & I(\bm x; \bm z) \label{IB_problem_lb1_a}\\
	\text{s.t.} \quad\; &  I({\bar {\bm x}}; \bm z) \leq C. \label{IB_problem_lb1_b}
	\end{align}
\end{subequations}
Note that since matrix $\bm H \bm H^H + \sigma^2 \bm I_K$ in (\ref{F}) is always invertible, the results obtained in this subsection always hold no matter $K \leq M$ or $K > M$.

To evaluate the bottleneck rate $I(\bm x; \bm z)$, we define an auxiliary Gaussian vector ${\bar {\bm x}}_g \sim {\cal {CN}} \left(\bm 0, {\mathbb E} \left[ {\bar {\bm x}} {\bar {\bm x}}^H \right] \right)$, let ${\bar {\bm z}}_g$ denote its representation, and choose $p(\bm z| {\bar {\bm x}})$ as well as $p({\bar {\bm z}}_g| {\bar {\bm x}}_g)$ to be conditionally Gaussian distribution, i.e.,
\begin{align}\label{z_zg}
\bm z & = {\bar {\bm x}} + \bm q, \nonumber\\
{\bar {\bm z}}_g & = {\bar {\bm x}}_g + \bm q,
\end{align}
where $\bm q \sim {\cal {CN}} (\bm 0, D \bm I_K)$ is independent of everything else.
Let
\begin{align}\label{mutual_yg_zg_lb1}
I({\bar {\bm x}}_g; {\bar {\bm z}}_g) & = \log \det \left( \bm I_K + \frac{{\mathbb E} \left[ {\bar {\bm x}} {\bar {\bm x}}^H \right]}{D} \right) \nonumber\\
& = C.
\end{align}
Then, rate $I({\bar {\bm x}}_g; {\bar {\bm z}}_g)$ is achievable and $D$ can be calculated from (\ref{mutual_yg_zg_lb1}).
Since $I({\bar {\bm x}}; \bm z) \leq I({\bar {\bm x}}_g; {\bar {\bm z}}_g)$, $I({\bar {\bm x}}; \bm z)$ is thus also achievable.

In the following, we obtain a lower bound to $I(\bm x; \bm z)$ by evaluating $h(\bm z | \bm H)$ and $h(\bm z | \bm x)$ separately, and then using 
\begin{align}\label{mutual_x_z2}
I(\bm x; \bm z) = & h(\bm z) - h(\bm z | \bm x) \nonumber\\
\geq & h(\bm z | \bm H) - h(\bm z | \bm x).
\end{align}
First, since $\bm z$ is conditionally Gaussian given $\bm H$, we have
\begin{equation}\label{diff_z_H_lb1}
h(\bm z | \bm H) \!=\! {\mathbb E} \left[ \log (\pi e)^K \det \left( \bm F^H \bm H \bm H^H \bm F \!+\! \sigma^2 \bm F^H \bm F \!+\! D \bm I_K \right) \right]\!.
\end{equation}
Next, using the fact that conditioning reduces entropy and Gaussian distribution maximizes the entropy over all distributions with the same variance \cite[Theorem 8.6.5]{cover2012elements}, we have
\begin{align}\label{diff_z_x_lb1}
h(\bm z | \bm x) & = h\left( \bm z - {\mathbb E} (\bm z | \bm x) | \bm x \right)\nonumber\\
& = h \left( \left( \bm F^H \bm H - {\mathbb E} \left[\bm F^H \bm H\right] \right) \bm x + \bm F^H \bm n + \bm q | \bm x  \right)\nonumber\\
& \leq h \left( \left( \bm F^H \bm H - {\mathbb E} \left[\bm F^H \bm H\right] \right) \bm x + \bm F^H \bm n + \bm q \right)\nonumber\\
& \leq \log (\pi e)^K \det(\bm G),
\end{align}
where 
\begin{align}\label{G}
\bm G & = {\mathbb E} \left[ \left( \bm F^H \bm H - {\mathbb E} \left[\bm F^H \bm H\right] \right) \left( \bm H^H \bm F - {\mathbb E} \left[\bm H^H \bm F\right] \right)\right.\nonumber\\
& \quad \left. + \sigma^2 \bm F^H \bm F \right] + D \bm I_K\nonumber\\
& = {\mathbb E} \left[ \bm F^H \bm H \bm H^H \bm F \right] - {\mathbb E} \left[ \bm F^H \bm H \right]  {\mathbb E} \left[ \bm H^H \bm F \right]\nonumber\\
& \quad + \sigma^2 {\mathbb E} \left[ \bm F^H \bm F \right] + D \bm I_K.
\end{align}
Combining (\ref{mutual_x_z2}), (\ref{diff_z_H_lb1}), and (\ref{diff_z_x_lb1}), we can get a lower bound to $I(\bm x; \bm z)$ as shown in the following theorem.

\begin{theorem}\label{theorem_MMSE}
	With MMSE estimate at the relay, a lower bound to $I(\bm x; \bm z)$ can be obtained as follows 
	\begin{align}\label{R_lb4}
	\!\!\!\! R^{\text {lb}2} & = T {\mathbb E} \left[ \log \left( \frac{\lambda}{\lambda + \sigma^2} + D \right) \right] + (K-T) \log D\nonumber\\
	\!\!\!\! & - K \log \left\{ \frac{T}{K} {\mathbb E} \left[ \frac{\lambda}{\lambda + \sigma^2} \right] - \frac{T^2}{K^2} \left( {\mathbb E} \left[ \frac{\lambda}{\lambda + \sigma^2} \right] \right)^2 + D \right\},
	\end{align}
	where 
	\begin{equation}\label{D_2}
	D = \frac{\frac{T}{K} {\mathbb E} \left[ \frac{\lambda}{\lambda + \sigma^2} \right]}{2^{\frac{C}{K}} - 1},
	\end{equation}
	and the expectations can be calculated by using the pdf of $\lambda$ in (\ref{pdf}).
\end{theorem}

\itshape \textbf{Proof:}  \upshape
	See Appendix \ref{prove_theorem_MMSE}.
\hfill $\Box$

\begin{lemma}\label{lemma_MMSE}
	When $M \rightarrow + \infty$ or when $K \leq M$ and $\rho \rightarrow + \infty$, lower bound $R^{\text {lb}2}$ tends asymptotically to $C$.
	When $K \leq M$ and $C \rightarrow + \infty$, 
	\begin{align}\label{R_lb4_appro}
	R^{\text {lb}2} & \rightarrow K {\mathbb E} \left[ \log \left( \frac{\lambda}{\lambda + \sigma^2} \right) \right]\nonumber\\
	& - K \log \left\{ {\mathbb E} \left[ \frac{\lambda}{\lambda + \sigma^2} \right] - \left( {\mathbb E} \left[ \frac{\lambda}{\lambda + \sigma^2} \right] \right)^2 \right\}.
	\end{align}
\end{lemma}

\itshape \textbf{Proof:}  \upshape
	See Appendix \ref{prove_lemma_MMSE}.
\hfill $\Box$

\section{Numerical Results}
\label{simulation}

In this section, we investigate the lower bounds obtained by the proposed achievable schemes and compare them with the upper bound derived in Section \ref{informed_ub}.
When performing the QCI scheme, we choose the quantization levels as quantiles for the sake of convenience.

\begin{figure}
	\centering
	\includegraphics[scale=0.50]{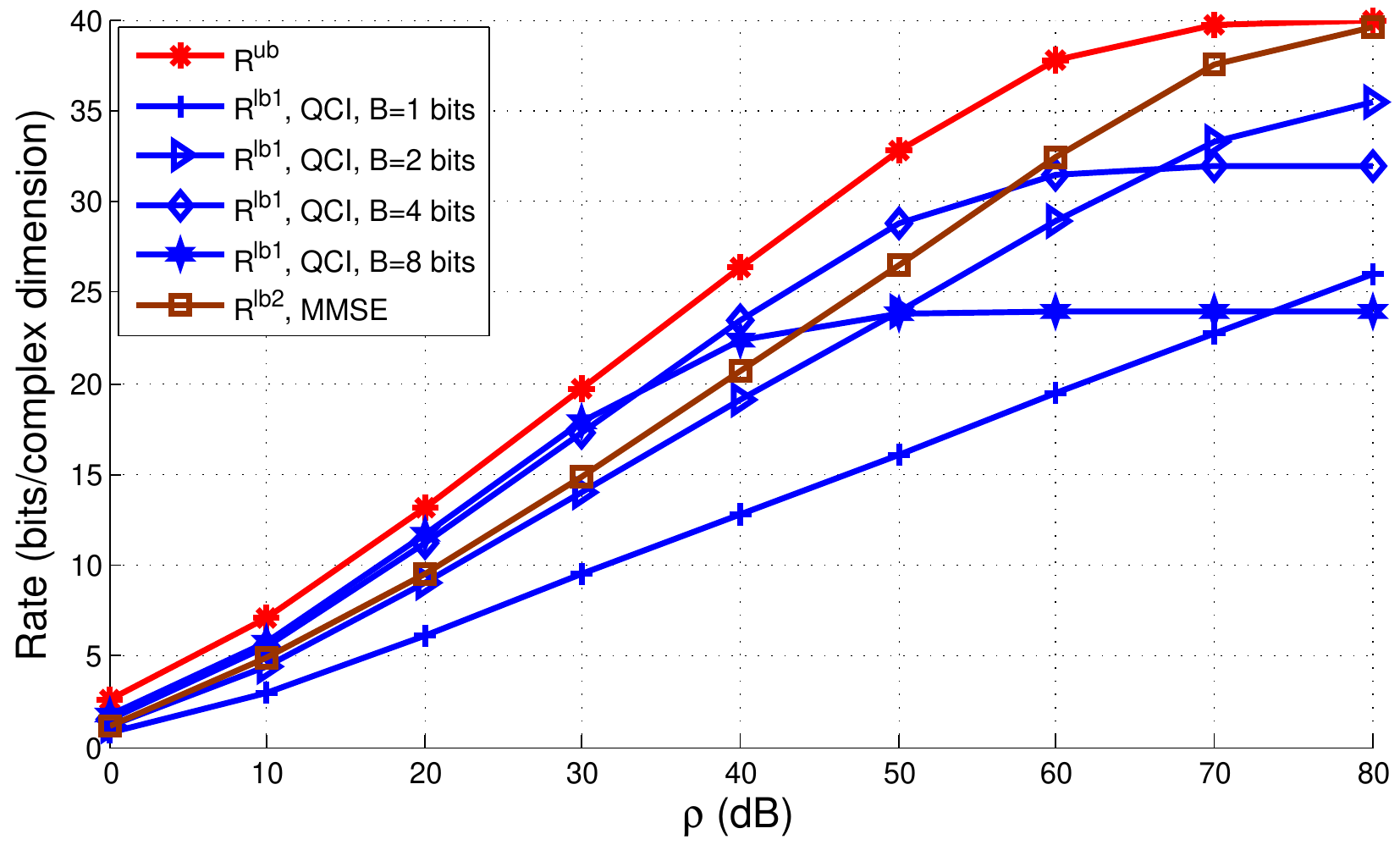}
	\caption{Upper and lower bounds to the bottleneck rate versus $\rho$ with $K=M=2$ and $C = 40$ bits/complex dimension.}
	\label{Fig1}
\end{figure}

\begin{figure}
	\centering
	\includegraphics[scale=0.50]{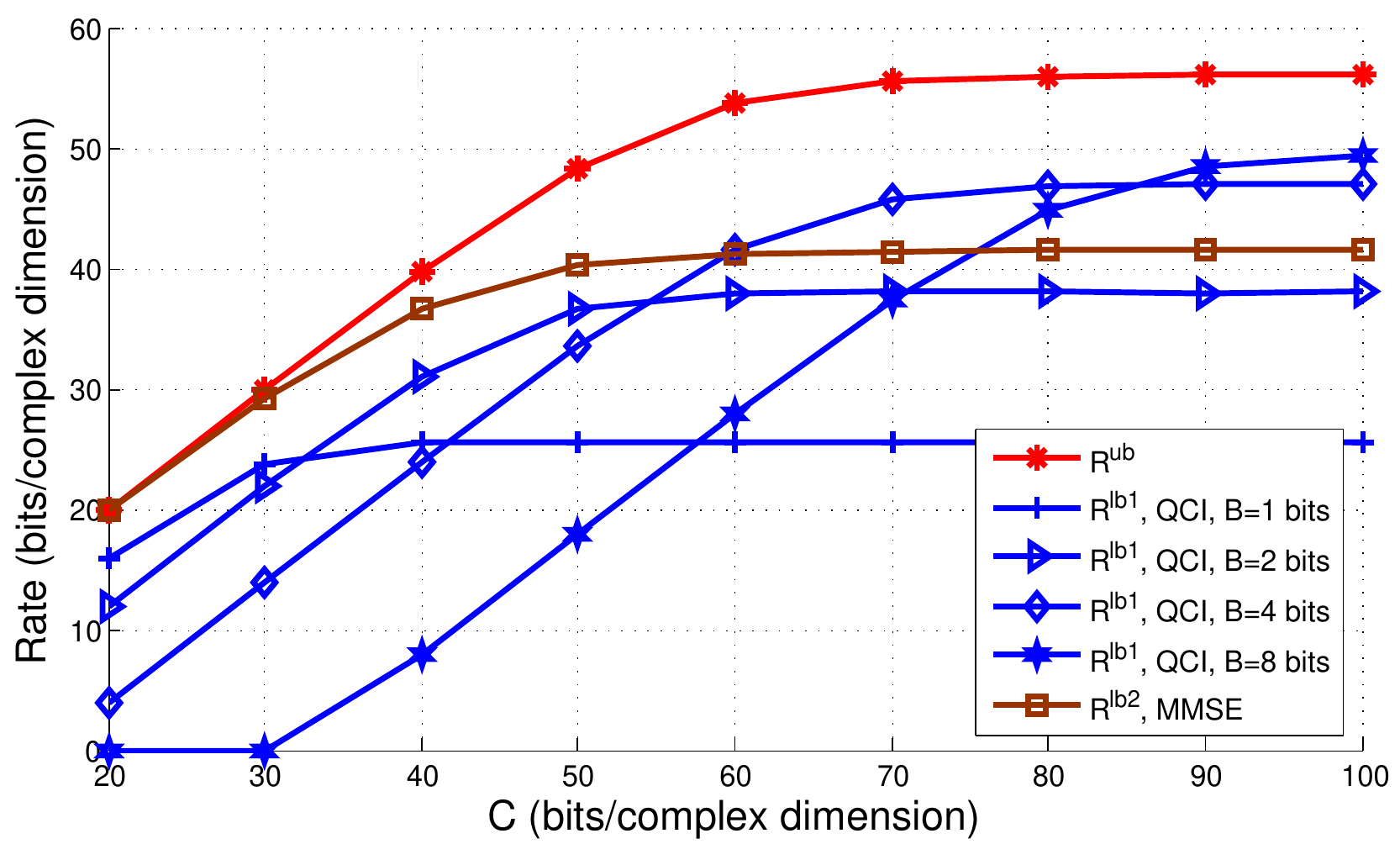}
	\caption{Upper and lower bounds to the bottleneck rate versus $C$ with $K=M=4$ and $\rho=40$dB.}
	\label{Fig2}
\end{figure}

In Fig.~\ref{Fig1}, the upper and lower bounds are depicted versus SNR $\rho$.
It can be found that when $\rho$ is small and $4$ or $8$ bits are applied to quantize the noise levels, the QCI scheme outperforms the MMSE scheme.
As $\rho$ grows large, $R^{\text {lb}2}$ obtained by the MMSE scheme approaches $C$ and is larger than $R^{\text {lb}1}$.
This is because when $\rho$ is small, the bottleneck rate is mainly limited by the capacity of channel~1, and the QCI scheme works well in this case since partial CSI, i.e., the noise level of each sub-channel, is conveyed to the destination node.
When $\rho$ is large, the MMSE scheme can get an accurate estimate and it does not require CSI feedback.
The MMSE scheme thus performs better when $\rho$ is large.

The effect of the bottleneck constraint $C$ is investigated in Fig.~\ref{Fig2}.
It can be found that as $C$ increases, all bounds grow and converge to different constants, which can be calculated based on Lemma \ref{lemma_ub}, Lemma \ref{lemma_QCI}, and Lemma \ref{lemma_MMSE}, respectively.
Fig.~\ref{Fig2} also shows $R^{\text {lb}2}$ virtually achieves the upper bound when $C$ is small, while when $C$ is large, the QCI scheme outperforms the MMSE scheme thanks to CSI feedback.

\begin{figure}
	\centering
	\includegraphics[scale=0.50]{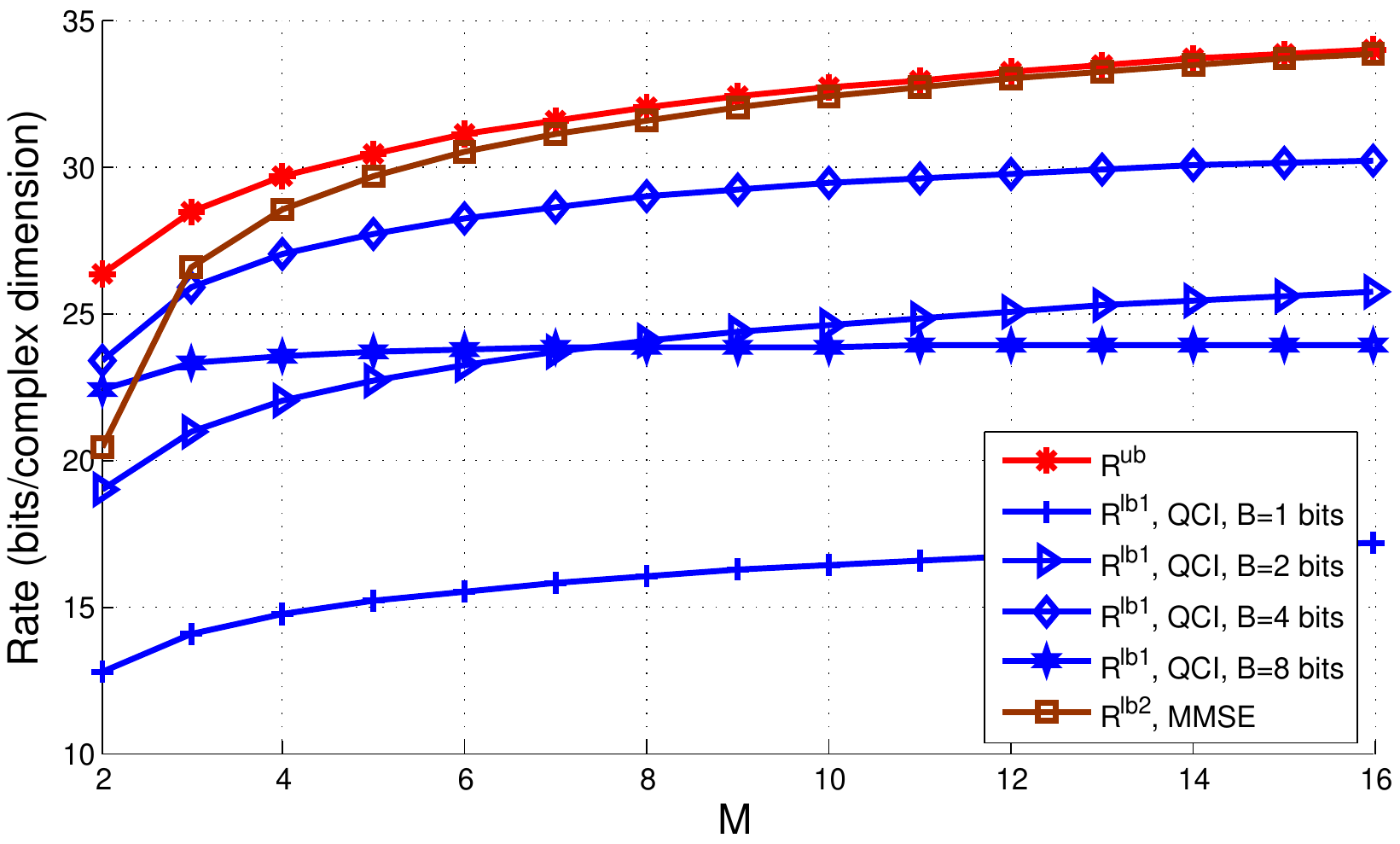}
	\caption{Upper and lower bounds to the bottleneck rate versus $M$ with $K=2$, $\rho=40$dB, and $C = 40$ bits/complex dimension.}
	\label{Fig3}
\end{figure}

Fig.~\ref{Fig3} depicts the bounds versus the number of relay antennas $M$.
As $M$ increases, $R^{\text {lb}2}$ quickly approaches $R^{\text {ub}}$.
It is also shown that the result for the limit case in Lemma \ref{lemma_QCI}, i.e., when $M \rightarrow + \infty$, we can always find suitable quantization points ${\cal B} = \{ b_1, \cdots, b_J \}$ such that $R^{\text {lb}1} \rightarrow C$, does not hold here.
This is because when performing the QCI scheme, we choose the quantization levels as quantiles.
The choice of quantization points ${\cal B} = \{ b_1, \cdots, b_J \}$ is thus restricted.

\section{Conclusions}
\label{conclusion}

This work extends the IB problem of the scalar case in \cite{caire2018information} to the case of MIMO Rayleigh fading channels.
Due to the information bottleneck constraint, the destination node cannot get the perfect CSI from the relay.
Our results show that with simple symbol-by-symbol oblivious relay processing and compression, we can get bottleneck rate close to the upper bound on a wide range of relevant system parameters.

%%%%%%%%%%%%%%%%%%%%%%%%%%%%%%%%%%%%%%%%%%%%%%%%%%%%%%%%%%%%%%%%%
\appendices

\section{Proof of Theorem \ref{theorem_ub}}
\label{prove_theorem_ub}

Before proving Theorem \ref{theorem_ub}, we first consider the following scalar Gaussian channel
\begin{equation}\label{scalar_channel}
y = s x + n,
\end{equation}
where $x \sim {\cal CN} (0, 1)$, $n \sim {\cal CN} (0, \sigma^2)$, and $s \in {\mathbb C}$ is the deterministic channel gain.
With bottleneck constraint $C$, the IB problem for (\ref{scalar_channel}) has been studied in \cite{winkelbauer2014rate} and the optimal bottleneck rate is given by 
\begin{equation}\label{bottleneck_rate}
R_0 = \log\left(1 + \rho |s|^2\right) - \log\left(1 + \rho |s|^2 2^{-C} \right).
\end{equation}
In the following, we show that (\ref{IB_problem1}) can be decomposed into a set of parallel scalar IB problems, and (\ref{bottleneck_rate}) can then be applied to get upper bound $R^{\text {ub}}$ in Theorem \ref{theorem_ub}.

According to the definition of conditional entropy, problem (\ref{IB_problem1}) can be rewritten as
\begin{subequations}\label{IB_problem2}
	\begin{align}
	\mathop {\max }\limits_{p(\bm z| \bm y, \bm H)} \quad & \int I(\bm x; \bm z| \bm H = {\mathbb H}) p_{\bm H} (\mathbb H) d \bm H \label{IB_problem2_a}\\
	\text{s.t.} \quad\;\;\; &  \int I(\bm y; \bm z| \bm H = {\mathbb H}) p_{\bm H} (\mathbb H) d \bm H \leq C, \label{IB_problem2_b}
	\end{align}
\end{subequations}
where $\mathbb H$ is a realization of $\bm H$.
Let $\bm U \bm \varLambda \bm U^H$ denote the eigendecomposition of $\bm H \bm H^H$, where $\bm U$ is a unitary matrix whose columns are the eigenvectors of $\bm H \bm H^H$, and $\bm \varLambda$ is a diagonal matrix whose diagonal elements are the eigenvalues of $\bm H \bm H^H$.
Since the rank of $\bm H \bm H^H$ is no greater than $T = \min \{K, M\}$, there are at most $T$ positive diagonal entries in $\bm \varLambda$.
Denote them by $\lambda_t$, where $t \in {\cal T}$ and ${\cal T} = \{1, \cdots, T\}$.
Let 
\begin{align}\label{hat_y}
{\hat {\bm y}} & = \bm U^H \bm y\nonumber\\
& = \bm U^H \bm H \bm x + \bm U^H \bm n.
\end{align}
Then, for a given channel realization $\bm H = \mathbb H$, ${\hat {\bm y}}$ is conditionally Gaussian, i.e.,
\begin{equation}\label{hat_y_distr}
{\hat {\bm y}} | \bm H = \mathbb H \sim {\cal CN} (\bm 0, \bm \varLambda + \sigma^2 \bm I_M).
\end{equation} 
Since 
\begin{equation}
I(\bm x; \bm y| \bm H = {\mathbb H}) = I(\bm x; {\hat {\bm y}}| \bm H = {\mathbb H}),
\end{equation}
we work with ${\hat {\bm y}}$ instead of $\bm y$ in the following.

Based on (\ref{IB_problem2}) and (\ref{hat_y_distr}), it is known that MIMO channel $p({\hat {\bm y}}| \bm x, \bm H)$ can be first divided into a set of parallel channels for different realizations of $\bm H$, and each channel $p({\hat {\bm y}}| \bm x, \bm H = \mathbb H)$ can be further divided into $T$ independent scalar Gaussian channels with SNRs $ \rho \lambda_t, \forall t \in {\cal T}$.
Accordingly, problem (\ref{IB_problem1}) can be decomposed into a set of parallel IB problems.
For a scalar Gaussian channel with SNR $ \rho \lambda_t$, let $c_t^{\text {ub}}$ denote the allocation of the bottleneck constraint $C$ and $R_t^{\text {ub}}$ denote the corresponding rate.
According to (\ref{bottleneck_rate}), we have
\begin{equation}\label{bottleneck_rate1}
R_t^{\text {ub}} =  \log\left(1 + \rho \lambda_t\right) - \log\left(1 + \rho \lambda_t 2^{-c_t^{\text {ub}}}\right).
\end{equation}
Then, the solution of problem (\ref{IB_problem1}) can be obtained by solving the following problem
\begin{subequations}\label{IB_problem3}
	\begin{align}
	\mathop {\max }\limits_{\{c_t^{\text {ub}}\}} \quad & \sum_{t=1}^{T} {\mathbb E} \left[ R_t^{\text {ub}} \right] \label{IB_problem3_a}\\
	\text{s.t.} \quad\; &  \sum_{t=1}^{T} {\mathbb E} \left[ c_t^{\text {ub}} \right] \leq C. \label{IB_problem3_b}
	\end{align}
\end{subequations}
Assume that $\lambda_t, \forall t \in {\cal T}$ are unordered positive eigenvalues of $\bm H \bm H^H$. \footnote{Note that when deriving the upper and lower bounds in this paper, we consider the unordered positive eigenvalues of $\bm H \bm H^H$ or $\bm H^H \bm H$ since it simplifies the analysis.
If the ordered positive eigenvalues of $\bm H \bm H^H$ or $\bm H^H \bm H$ are considered, it can be readily proven by following similar steps in \cite[Subsetion 4.2]{telatar1999capacity} that we arrive at problems equivalent to those in this paper.}
Then, they follow the same distribution.
For convenience, define a new variable $\lambda$ which follows the same distribution as $\lambda_t$.
The subscript `$t$' in $c_t^{\text {ub}}$ and $R_t^{\text {ub}}$ can thus be omitted.
In order to distinguish from $R^{\text {ub}}$ in (\ref{R_up_KM}), we use $R_0^{\text {ub}}$ to denote the bottleneck rate corresponding to $c^{\text {ub}}$, i.e.,
\begin{equation}\label{bottleneck_rate0}
R_0^{\text {ub}} = \log\left(1 + \rho \lambda\right) - \log\left(1 + \rho \lambda 2^{-c^{\text {ub}}}\right).
\end{equation}
Then, we have
\begin{align}
&\sum_{t=1}^{T} {\mathbb E} \left[ R_t^{\text {ub}} \right] = T {\mathbb E} \left[ R_0^{\text {ub}} \right],\nonumber\\
&\sum_{t=1}^{T} {\mathbb E} \left[ c_t^{\text {ub}} \right] = T {\mathbb E} \left[ c^{\text {ub}} \right].
\end{align}
Problem (\ref{IB_problem3}) thus becomes
\begin{subequations}\label{IB_problem4}
	\begin{align}
	\mathop {\max }\limits_{c^{\text {ub}}} \quad & {\mathbb E} \left[ R_0^{\text {ub}} \right] \label{IB_problem4_a}\\
	\text{s.t.} \quad\; &  {\mathbb E} \left[ c^{\text {ub}} \right] \leq \frac{C}{T}. \label{IB_problem4_b}
	\end{align}
\end{subequations}
This problem can be solved by the water-filling method.
Consider the Lagrangian
\begin{equation}\label{Lagrangian}
{\cal L} = {\mathbb E} \left[ - R_0^{\text {ub}} + \alpha c^{\text {ub}} \right] - \frac{\alpha C}{T},
\end{equation}
where $\alpha$ is the Lagrange multiplier.
The KKT condition for the optimality is
\begin{equation}\label{KKT}
\frac{\partial \cal L}{ \partial c^{\text {ub}} } \left\{
\begin{array}{ll}
=0,&  {\text {if}}~ c^{\text {ub}} > 0\\
\leq 0,&  {\text {if}}~ c^{\text {ub}} = 0\\
\end{array} \right..
\end{equation} 
Then,
\begin{equation}\label{optimal_c}
c^{\text {ub}} = \left\{
\begin{array}{ll}
\log \frac{\rho \lambda}{ \nu},&  {\text {if}}~ \lambda > \frac{\nu}{\rho}\\
0,&  {\text {if}}~ \lambda \leq \frac{\nu}{\rho}\\
\end{array} \right.,
\end{equation} 
where $\nu = \alpha/(1-\alpha)$ and it is chosen such that the following bottleneck constraint is met
\begin{equation}\label{bottle_constr}
{\mathbb E} \left[ \log \frac{\rho \lambda}{ \nu} ~| \lambda > \frac{\nu}{\rho} \right] {\text {Pr}}\left\{ \lambda > \frac{ \nu}{\rho} \right\} = \frac{C}{T}.
\end{equation}
The informed receiver upper bound is thus given by
\begin{equation}\label{R_up}
R^{\text {ub}} = T {\mathbb E} \left[ \log \left(1 + \rho \lambda\right) - \log (1 + \nu) ~| \lambda > \frac{\nu}{\rho} \right] {\text {Pr}}\left\{ \lambda > \frac{\nu}{\rho} \right\}.
\end{equation}

From the definition of $\bm H$ in (\ref{relay_obser}), it is known that when $K \leq M$ (resp., when $K > M$), $\bm H^H \bm H$ (resp., $\bm H \bm H^H$) is a central complex Wishart matrix with $M$ (resp., $K$) degrees of freedom and covariance matrix $\bm I_K$ (resp., $\bm I_M$), i.e., $\bm H^H \bm H \sim {{\cal {CW}}_K} (M, \bm I_K)$ (resp., $\bm H \bm H^H \sim {{\cal {CW}}_M} (K, \bm I_M)$) \cite{tulino2004random}.
Since $\lambda$ can be seen as one of the unordered positive eigenvalues of $\bm H^H \bm H$ or $\bm H \bm H^H$, its pdf is thus given by \cite[Theorem 2.17]{tulino2004random}, \cite{telatar1999capacity} 
\begin{equation}\label{pdf}
f_\lambda (\lambda) = \frac{1}{T} \sum_{i=0}^{T-1} \frac{i!}{(i+S-T)!} \left[ L_i^{S-T} (\lambda) \right]^2 \lambda^{S-T} e^{-\lambda},
\end{equation}
where $S = \max \{K, M\}$ and the Laguerre polynomials are
\begin{equation}\label{Laguerre}
L_i^{S-T} (\lambda) = \frac{e^{\lambda}}{i! \lambda^{S-T}} \frac{d^i}{d \lambda^i} \left( e^{-\lambda} \lambda^{S-T+i} \right).
\end{equation}
Substituting (\ref{pdf}) and (\ref{Laguerre}) into (\ref{R_up}) and (\ref{bottle_constr}), (\ref{R_up_KM}) and (\ref{bottle_constr_KM}) can be obtained.
Theorem \ref{theorem_ub} is thus proven.

\section{Proof of Lemma \ref{lemma_ub}}
\label{prove_lemma_ub}

In order to prove that $R^{\text {ub}}$ approaches $C$ as $M \rightarrow + \infty$, we first look at the special case with $K=1$.
In this case, $S=M$ and $T=1$.
From (\ref{Laguerre}) and (\ref{pdf}), we have $L_0^{S-T}=1$ and the pdf of $\lambda$
\begin{equation}\label{pdf_1M}
f_\lambda (\lambda) = \frac{\lambda^{M-1} e^{-\lambda}}{(M-1)!},
\end{equation}
which shows that $\lambda$ follows Erlang distribution with shape parameter $M$ and rate parameter $1$, i.e., $ \lambda \sim {\text {Erlang}}(M,1)$.
The expectation of $\lambda$ is thus $M$.
As $M \rightarrow + \infty$, $f_\lambda (\lambda)$ becomes a delta function \cite{lee1990estimate}.
Hence, for a sufficiently small positive real number $\epsilon$, 
\begin{align}\label{impulse}
& \lim_{M \rightarrow + \infty} {\text {Pr}}\left\{ |\lambda - M| \leq \epsilon \right\} \rightarrow 1,\nonumber\\
& \lim_{M \rightarrow + \infty} {\text {Pr}}\left\{ |\lambda - M| > \epsilon \right\} \rightarrow 0.
\end{align}
Then, when $M \rightarrow + \infty$, the bottleneck constraint (\ref{bottle_constr_KM})
\begin{align}
\int_{\frac{\nu}{\rho}}^{\infty} \left( \log \frac{\rho \lambda}{ \nu} \right) f_\lambda (\lambda) d \lambda & = C \nonumber\\
& \rightarrow \int_{M - \epsilon}^{M + \epsilon} \left( \log \frac{\rho \lambda}{ \nu} \right) f_\lambda (\lambda) d \lambda \nonumber\\
& \rightarrow \log \frac{\rho M}{ \nu},
\end{align}
based on which we get
\begin{equation}\label{nu1}
\frac{\nu}{M} \rightarrow \rho 2^{-C}.
\end{equation}
Using (\ref{R_up_KM}), (\ref{impulse}), and (\ref{nu1}), it is known that when $M \rightarrow + \infty$,
\begin{align}
R^{\text {ub}} = & \int_{\frac{\nu}{\rho}}^{\infty} \left[ \log \left(1 + \rho \lambda\right) - \log (1 + \nu)\right] f_\lambda (\lambda) d \lambda \nonumber\\
\rightarrow & \int_{M - \epsilon}^{M + \epsilon} \left( \log \frac{1 + \rho \lambda}{1 + \nu} \right) f_\lambda (\lambda) d \lambda \nonumber\\
\rightarrow & \log \frac{1 + \rho M}{1 + \nu}\nonumber\\
\rightarrow & C.
\end{align}

Next, we consider the general case.
For any positive integer $K$, when $M\rightarrow + \infty$, based on the definition of $\bm H$ and the strong law of large numbers, we almost surely have $\bm H^H \bm H - M \bm I_K \rightarrow \bm 0$.
Since $\bm H \bm H^H$ and $\bm H^H \bm H$ have the same positive eigenvalues, $\lambda - M \rightarrow 0$ almost surely.
(\ref{impulse}) thus also holds for this general case.
Then, 
\begin{align}
\int_{\frac{\nu}{\rho}}^{\infty} \left( \log \frac{\rho \lambda}{ \nu} \right) f_\lambda (\lambda) d \lambda & = \frac{C}{T} \nonumber\\
& \rightarrow \int_{M - \epsilon}^{M + \epsilon} \left( \log \frac{\rho \lambda}{ \nu} \right) f_\lambda (\lambda) d \lambda \nonumber\\
& \rightarrow \log \frac{\rho M}{ \nu},
\end{align}
based on which we get
\begin{equation}
\frac{\nu}{M} \rightarrow \rho 2^{-C/T}.
\end{equation}
Hence, when $M\rightarrow + \infty$,
\begin{align}
R^{\text {ub}} \rightarrow & T \int_{\frac{\nu}{\rho}}^{\infty} \left[ \log \left(1 + \rho \lambda\right) - \log (1 + \nu)\right] f_\lambda (\lambda) d \lambda \nonumber\\
\rightarrow & T \int_{M - \epsilon}^{M + \epsilon} \left( \log \frac{1 + \rho \lambda}{1 + \nu} \right) f_\lambda (\lambda) d \lambda \nonumber\\
\rightarrow & T \log \frac{1 + \rho M}{1 + \nu}\nonumber\\
\rightarrow & C.
\end{align}

Now we prove that $R^{\text {ub}}$ approaches $C$ as $\rho \rightarrow + \infty$.
From (\ref{bottle_constr_KM}), it can be seen that $\int_{\frac{\nu}{\rho}}^{\infty} \left( \log \frac{\rho \lambda}{ \nu} \right) f_\lambda (\lambda) d \lambda$ reduces with $\nu$.
Therefore, when $\rho \rightarrow + \infty$, to ensure that constraint (\ref{bottle_constr_KM}) holds, $\nu$ becomes large.
Then, we have
\begin{align}
R^{\text {ub}} & = T \int_{\frac{\nu}{\rho}}^{\infty} \left[ \log \left(1 + \rho \lambda\right) - \log (1 + \nu)\right] f_\lambda (\lambda) d \lambda\nonumber\\
& \rightarrow T \int_{\frac{\nu}{\rho}}^{\infty} \left[ \log \left(\rho \lambda \right) - \log \nu \right] f_\lambda (\lambda) d \lambda\nonumber\\
& = C.
\end{align}

In addition, when $C \rightarrow + \infty$, it can be found from (\ref{bottle_constr_KM}) that $\nu \rightarrow 0$. 
Using (\ref{R_up_KM}), we can get (\ref{R_up_appro}), which is the capacity of Channel~1.
This completes the proof.

\section{Proof of Theorem \ref{theorem_QCI}}
\label{prove_theorem_QCI}

Since ${\hat {\bm n}}_g \sim {\cal CN} \left( \bm 0,  \bm A_1'\right)$ and $\ceil[\big]{a_k}_{\cal B}$ has $J$ possible values, i.e., $b_1, \cdots, b_J$, the channel in (\ref{x_bar}) can be divided into $KJ$ independent scalar Gaussian sub-channels with noise power $\ceil[\big]{a_k}_{\cal B} = b_j$ for each sub-channel.
For the sub-channel with noise power $\ceil[\big]{a_k}_{\cal B} = b_j$, let $c_{k,j}$ denote the allocation of the bottleneck constraint $C$ and $R_{k,j}$ denote the corresponding rate.
According to (\ref{bottleneck_rate}), we have
\begin{equation}\label{bottleneck_rate2}
R_{k,j} =  \log\left(1 + \rho_j\right) - \log\left(1 + \rho_j 2^{-c_{k,j}}\right),
\end{equation}
where $\rho_j = \frac{1}{b_j}$.
Since $b_J = + \infty$, we let $R_{k,J} = 0$ and $c_{k,J} = 0$.
Note that based on \cite[(16)]{winkelbauer2014rate}, the representation of ${\hat {\bm x}}_g$, i.e., ${\hat {\bm z}}_g$, can be constructed by adding independent fading and Gaussian noise to each element of ${\hat {\bm x}}_g$ in (\ref{x_bar}).
Denote 
\begin{equation}\label{prob}
P_{k,j} = {\text {Pr}}\left\{ \ceil[\big]{a_k}_{\cal B} = b_j \right\}.
\end{equation}
Then, the optimal $I(\bm x; {\hat {\bm z}}_g| \bm A_1')$ is equal to the objective function of the following problem
\begin{subequations}\label{IB_problem5}
\begin{align}
\mathop {\max }\limits_{\{c_{k,j}\}} \quad & \sum_{k=1}^K \sum_{j=1}^{J-1} P_{k,j} R_{k,j} \label{IB_problem5_a}\\
\text{s.t.} \quad\; &  \sum_{k=1}^K \sum_{j=1}^{J-1} P_{k,j} c_{k,j} \leq C - \sum_{k=1}^K H_k, \label{IB_problem5_b}
\end{align}
\end{subequations}
where $H_k = - \sum_{j=1}^J P_{k,j} \log P_{k,j}$.

Since $K \leq M$, as stated in Appendix \ref{prove_theorem_ub}, $\bm H^H \bm H \sim {{\cal {CW}}_K} (M, \bm I_K)$.
Then, $(\bm H^H \bm H)^{-1}$ follows a complex inverse Wishart distribution and the diagonal elements of $(\bm H^H \bm H)^{-1}$ are identically inverse chi squared distributed with $M-K+1$ degrees of freedom \cite{brennan1982adaptive}.
Let $\eta$ denote one of the diagonal element of $(\bm H^H \bm H)^{-1}$.
The pdf of $\eta$ is thus given by
\begin{equation}\label{pdf_eta}
f_\eta (\eta) = \frac{2^{-(M-K+1)/2}}{\Gamma \left( \frac{M-K+1}{2} \right)} \eta^{-(M-K+1)/2-1} e^{-1/(2 \eta)}.
\end{equation}
Since $\bm A = \sigma^2 (\bm H^H \bm H)^{-1}$, the diagonal entries of $\bm A$, i.e., $a_k, \forall k \in {\cal K}$, are marginally identically distributed.
Let $a$ denote a new variable which has the same distribution as $a_k$.
$a$ thus follows the same distribution as $\sigma^2 \eta$ and its pdf is given by
\begin{align}\label{pdf_a}
f_a (a) & = \frac{1}{\sigma^2} f_\eta \left( \frac{a}{\sigma^2} \right) \nonumber\\
& = \frac{(2/\sigma^2)^{-(M-K+1)/2}}{ \Gamma \left( \frac{M-K+1}{2} \right)} a^{-(M-K+1)/2-1} e^{-\sigma^2/(2 a)}.
\end{align}
In addition, $P_{k,j}$, $R_{k,j}$, and $c_{k,j}$ can be simplified to $P_j$, $R_j$, and $c_j$ by dropping subscript `$k$'.
Using (\ref{pdf_a}), pmf $P_j$ can be calculated as follows
\begin{align}\label{P_j}
P_j & = {\text {Pr}}\left\{ \ceil[\big]{a}_{\cal B} = b_j \right\} \nonumber\\
& = {\text {Pr}}\left\{ b_{j-1} < a \leq b_j \right\} \nonumber\\
& = \int_{b_{j-1}}^{b_j} f_a (a) d a.
\end{align}
Problem (\ref{IB_problem5}) thus becomes
\begin{subequations}\label{IB_problem6}
	\begin{align}
	\mathop {\max }\limits_{\{c_j\}} \quad & \sum_{j=1}^{J-1} K P_j R_j \label{IB_problem6_a}\\
	\text{s.t.} \quad\; &  \sum_{j=1}^{J-1} K P_j c_j \leq C - K H_0, \label{IB_problem6_b}
	\end{align}
\end{subequations}
where 
\begin{align}\label{RPH}
& R_j =  \log\left(1 + \rho_j\right) - \log\left(1 + \rho_j 2^{-c_j}\right), \nonumber\\
& H_0 = - \sum_{j=1}^J P_j \log P_j.
\end{align}
Analogous to problem (\ref{IB_problem4}), (\ref{IB_problem6}) can be optimally solved by the water-filling method.
The following lower bound to the bottleneck rate can thus be obtained
\begin{equation}
R^{\text {lb}1} = \sum_{j=1}^{J-1} K P_j \left[ \log \left(1 + \rho_j\right) - \log (1 + \rho_j 2^{-c_j} ) \right].
\end{equation}
where $c_j = \left[ \log \frac{\rho_j}{\nu} \right]^+$ and $\nu$ is chosen such that the bottleneck constraint 
\begin{equation}
\sum_{j=1}^{J-1} K P_j c_j = C - K H_0,
\end{equation}
is met.
Theorem \ref{theorem_QCI} is then proven.

\section{Proof of Lemma \ref{lemma_ineq_QCI}}
\label{prove_lemma_ineq_QCI}

Since $\bm \varPhi$ is a diagonal matrix with positive and real diagonal entries, it is invertible.
Denote
\begin{align}
\bm z' & = \bm \varPhi^{-1} \bm z \nonumber\\
& = \bm x + {\hat {\bm n}} + \bm \varPhi^{-1} {\hat {\bm n}}_g', \nonumber\\
{\hat {\bm z}}_g' & = \bm \varPhi^{-1} {\hat {\bm z}}_g \nonumber\\
& = \bm x + {\hat {\bm n}}_g + \bm \varPhi^{-1} {\hat {\bm n}}_g'.
\end{align}
For a given $\bm A_1'$, each element in ${\hat {\bm n}}$ is Gaussian distributed with zero mean and variance $\ceil[\big]{a_k}_{\cal B}$.
However, ${\hat {\bm n}}$ is not a Gaussian vector since $\bm H$ is unknown.
Hence, $\bm z'$ is not a Gaussian vector.
As for ${\hat {\bm z}}_g'$, from (\ref{x_bar}) and (\ref{z_bar}), it is known that ${\hat {\bm z}}_g' \sim {\cal CN}(\bm 0, \bm I_K + \bm A_1' + \bm \varPhi^{-2})$.

We first prove inequation (\ref{ineq3}).
\begin{align}\label{prove_ineq3}
& I({\hat {\bm x}}; \bm z| \bm A_1') \nonumber\\
= & I({\hat {\bm x}}; \bm z'| \bm A_1') \nonumber\\
= & h(\bm z'| \bm A_1') - h(\bm z'| {\hat {\bm x}}, \bm A_1') \nonumber\\
\overset{(a)}{\leq} & \mathbb{E} \left[ \log \det \left( \bm I_K + \mathbb{E} \left[ {\hat {\bm n}} {\hat {\bm n}}^H \right] + \bm \varPhi^{-2} \right) - \log \det \left( \bm \varPhi^{-2} \right) \right] \nonumber\\
\overset{(b)}{\leq} & \mathbb{E} \left[ \log \det \left( \bm I_K + \bm A_1' + \bm \varPhi^{-2} \right) - \log \det \left( \bm \varPhi^{-2} \right) \right] \nonumber\\
= & I({\hat {\bm x}}_g; {\hat {\bm z}}_g'| \bm A_1') \nonumber\\
= & I({\hat {\bm x}}_g; {\hat {\bm z}}_g| \bm A_1'),
\end{align}
where $(a)$ holds since Gaussian distribution maximizes the entropy over all distributions with the same variance, and $(b)$ follows by using Hadamard's inequality.

Denote $\bm x = (x_1, \cdots, x_K)^T$, $\bm z' = (z_1', \cdots, z_K')^T$, ${\hat {\bm z}}_g' = ({\hat z}_{g,1}', \cdots, {\hat z}_{g,K}')^T$, and $\bm \varPhi = {\text {diag}} \{\varphi_1, \cdots, \varphi_K\}$. 
Then, we prove inequation (\ref{ineq4}).
Using the chain rule of mutual information,
\begin{align}\label{prove_ineq4}
I(\bm x; \bm z| \bm A_1') = & I(\bm x; \bm z'| \bm A_1') \nonumber\\
\geq & \sum_{k=1}^K I(x_k; z_k'| \bm A_1') \nonumber\\
\overset{(a)}{=} & \sum_{k=1}^K I(x_k; {\hat z}_{g,k}'| \bm A_1') \nonumber\\
\overset{(b)}{=} & I(\bm x; {\hat {\bm z}}_g'| \bm A_1') \nonumber\\
= & I(\bm x; {\hat {\bm z}}_g| \bm A_1'),
\end{align}
where $(a)$ holds since for a given $\bm A_1'$, both $z_k'$ and ${\hat z}_{g,k}'$ follow ${\cal CN} \left( 0, 1 + \ceil[\big]{a_k}_{\cal B} + \varphi_k^{-2} \right)$, and $(b)$ follows since the elements in $\bm x$ and $\bm {\hat {\bm z}}_g'$ are independent.

Since $\bm \varPhi$ is optimally obtained when solving IB problem (\ref{problem_QCI}), bottleneck constraint (\ref{problem_QCI_b}) is thus satisfied and $I(\bm x; {\hat {\bm z}}_g| \bm A_1') = R^{\text {lb}1}$.
Then, from (\ref{prove_ineq3}) and (\ref{prove_ineq4}), we have
\begin{align}
I({\hat {\bm x}}; \bm z| \bm A_1') & \leq C - K H_0, \nonumber\\
I(\bm x; \bm z| \bm A_1') & \geq R^{\text {lb}1}.
\end{align}
This completes the proof.

\section{Proof of Lemma \ref{lemma_QCI}}
\label{prove_lemma_QCI}

As stated in Appendix \ref{prove_lemma_ub}, when $M \rightarrow + \infty$, $\bm H^H \bm H - M \bm I_K \rightarrow \bm 0$ almost surely.
Hence, $\bm A - \frac{\sigma^2}{M} \bm I_K \rightarrow \bm 0$.
Let $J=2$, $b_1 = \frac{\sigma^2}{M} + \epsilon$, and $b_2 = + \infty$, where $\epsilon$ is a sufficiently small positive real number.
Since $\bm A - \frac{\sigma^2}{M} \bm I_K \rightarrow \bm 0$, we have $P_1 \rightarrow 1$ and $H_0 \rightarrow 0$.
Then, from (\ref{R_lb1}) and (\ref{BT_constraint}), 
\begin{align}
c_1 & \rightarrow \frac{C}{K}, \nonumber\\
R^{\text {lb}1} & \rightarrow K \left[ \log \left(1 + \frac{M}{\sigma^2} \right) - \log \left(1 + \frac{M}{\sigma^2} 2^{-\frac{C}{K}} \right) \right] \nonumber\\
& \rightarrow C.
\end{align}

When $\rho \rightarrow + \infty$, $\sigma^2 \rightarrow 0$ and $\bm A \rightarrow \bm 0$.
By setting $J=2$ and $b_1$ small enough, it can be proven as above that $R^{\text {lb}1} \rightarrow C$.

When $C \rightarrow + \infty$, we could choose quantization points ${\cal B} = \{ b_1, \cdots, b_J \}$ with sufficiently large $J$ such that the diagonal entries of $\bm A_1$, which are continuously valued, can be represented precisely using the discretely valued points in ${\cal B}$, and the representation indexes of all diagonal entries can be transmitted to the destination node since $C$ is large enough.
On the other hand, as shown in (\ref{z_bar}), a representation of ${\hat {\bm x}}_g$ is
\begin{equation}\label{z_bar2}
{\hat {\bm z}}_g = \bm \varPhi {\hat {\bm x}}_g + {\hat {\bm n}}_g',
\end{equation}
where $\bm \varPhi$ is a diagonal matrix with positive and real diagonal entries, and ${\hat {\bm n}}_g' \sim {\cal CN} \left( \bm 0,  \bm I_K \right)$.
As $C \rightarrow + \infty$, according to \cite[(17) and (20)]{winkelbauer2014rate}, the diagonal entries of $\bm \varPhi$
\begin{align}\label{varphi_k}
\varphi_k & = \sqrt{\frac{\frac{1}{\ceil[\big]{a_k}_{\cal B}} + 2^C }{1 + \ceil[\big]{a_k}_{\cal B}} - \frac{1}{\ceil[\big]{a_k}_{\cal B}}} \nonumber\\
& \rightarrow \sqrt{\frac{2^C}{1 + \ceil[\big]{a_k}_{\cal B}}}, ~\forall~ k \in {\cal K}.
\end{align}
Since $\bm \varPhi$ is a diagonal matrix with positive and real diagonal entries, it is invertible.
Denote
\begin{align}\label{zg'}
{\hat {\bm z}}_g' & = \bm \varPhi^{-1} {\hat {\bm z}}_g \nonumber\\
& = {\hat {\bm x}}_g + \bm \varPhi^{-1} {\hat {\bm n}}_g'.
\end{align}
From (\ref{varphi_k}) it is known that the elements in noise vector $\bm \varPhi^{-1} {\hat {\bm n}}_g'$ have zero mean and very small (approaches $0$) power when $C \rightarrow + \infty$. 
Hence, $(\bm x, {\hat {\bm z}}_g') \to (\bm x, {\hat {\bm x}}_g)$ in distribution.
Then, based on \cite{csiszar1992arbitrarily}, we have
\begin{equation}\label{inequa2}
I(\bm x; {\hat {\bm x}}_g| \bm A_1') \leq \mathop {\lim \inf }\limits_{C \to + \infty} I(\bm x; {\hat {\bm z}}_g'| \bm A_1').
\end{equation}
In addition, since Gaussian noise vector ${\hat {\bm n}}_g$ (defined in (\ref{x_bar})) is independent of $\bm x$ and $\bm \varPhi^{-1} {\hat {\bm n}}_g'$ in (\ref{zg'}) is independent of both $\bm x$ and ${\hat {\bm n}}_g$, $\bm x \to {\hat {\bm x}}_g \to {\hat {\bm z}}_g'$ forms a Markov Chain.
Then, according to data-processing inequality, we have
\begin{equation}\label{inequa1}
I(\bm x; {\hat {\bm z}}_g'| \bm A_1') \leq I(\bm x; {\hat {\bm x}}_g| \bm A_1').
\end{equation}
Combining (\ref{inequa1}) and (\ref{inequa2}), we have 
\begin{equation}\label{inequa3}
I(\bm x; {\hat {\bm x}}_g| \bm A_1') \leq \mathop {\lim \inf }\limits_{C \to + \infty} I(\bm x; {\hat {\bm z}}_g'| \bm A_1') \leq I(\bm x; {\hat {\bm x}}_g| \bm A_1'),
\end{equation}
showing that the limit $\mathop {\lim \inf }\limits_{C \to + \infty} I(\bm x; {\hat {\bm z}}_g'| \bm A_1')$ exists and it is equal to $I(\bm x; {\hat {\bm x}}_g| \bm A_1')$.
Then, when $C \rightarrow + \infty$,
\begin{align}\label{R_lb1_C_infty}
R^{\text {lb}1} & = I(\bm x; {\hat {\bm z}}_g| \bm A_1') \nonumber\\
& = I(\bm x; {\hat {\bm z}}_g'| \bm A_1') \nonumber\\
& \rightarrow I(\bm x; {\hat {\bm x}}_g| \bm A_1') \nonumber\\
& = \mathbb{E} \left[ \log \det \left( \bm I_K + \bm A_1' \right) - \log \det \left( \bm A_1' \right) \right] \nonumber\\
& \rightarrow \mathbb{E} \left[ \log \det \left( \bm I_K + \bm A_1 \right) - \log \det \left( \bm A_1 \right) \right],
\end{align}

On the other hand, the capacity of Channel~1 is given by
\begin{align}\label{Capacity_channel_1}
I(\bm x; \bm y, \bm H) & = I(\bm x; \bm y| \bm H) \nonumber\\
& = \mathbb{E} \left[ \log \det \left( \bm H \bm H^H + \sigma^2 \bm I_M \right) - \log \det \left( \sigma^2 \bm I_M \right) \right] \nonumber\\
& = \mathbb{E} \left[ \log \det \left( \bm H^H \bm H + \sigma^2 \bm I_K \right) - \log \det \left( \sigma^2 \bm I_K \right) \right] \nonumber\\
& = \mathbb{E} \left[ \log \det \left( \bm I_K + \bm A \right) - \log \det \left( \bm A \right) \right].
\end{align}
To prove that (\ref{R_lb1_C_infty}) is upper bounded by (\ref{Capacity_channel_1}), we first give and prove the following lemma.
\begin{lemma}\label{mutuan_ine}
	For any $K$-dimensional positive definite matrix $\bm N$, let $\bm N_1 = \bm N \odot \bm I_K$, i.e., $\bm N_1$ consist of the diagonal elements of $\bm N$.
	Then,
	\begin{align}\label{ineq1}
	& \log \det \left( \bm I_K + \bm N \right) - \log \det \left( \bm N \right) \nonumber\\
	\geq & \log \det \left( \bm I_K + \bm N_1 \right) - \log \det \left( \bm N_1 \right).
	\end{align}
\end{lemma}

\itshape \textbf{Proof:}  \upshape
Obviously, (\ref{ineq1}) is equivalent to
\begin{align}\label{ineq2}
& \log \det \left( \bm N_1 \right) - \log \det \left( \bm N \right) \nonumber\\
\geq & \log \det \left( \bm I_K + \bm N_1 \right) - \log \det \left( \bm I_K + \bm N \right).
\end{align}
To prove (\ref{ineq2}), we introduce an auxiliary function $g_1 (x) = \log \det \left( x \bm I_K + \bm N_1 \right) - \log \det \left( x \bm I_K + \bm N \right)$ and show that $g_1 (x)$ decreases monotonically w.r.t. $x$ when $x \geq 0$.
By taking the first-order derivative to $g_1 (x)$, we have
\begin{equation}\label{derivative}
g_1'(x) = {\text {tr}} \left[ \left( x \bm I_K + \bm N_1 \right)^{-1} \right] - {\text {tr}} \left[ \left( x \bm I_K + \bm N \right)^{-1} \right].
\end{equation}
To prove $g_1'(x) \leq 0$, we show in the following that for any positive definite matrix $\bm O$, we always have
\begin{equation}\label{trace_inequa}
{\text {tr}} \left( {\bm O}_1^{-1} \right) \leq {\text {tr}} \left( \bm O^{-1} \right),
\end{equation}
where ${\bm O}_1$ consists of the diagonal elements of $\bm O$, i.e., ${\bm O}_1 = \bm O \odot \bm I_K$.
Denote the diagonal entries of $\bm O$ (or ${\bm O}_1$) by $\bm o = ( o_1, \cdots, o_K )^T$ and the eigenvalues of $\bm O$ by $\bm \theta = ( \theta_1, \cdots, \theta_K )^T$.
Since $\bm O$ is a positive definite matrix, the entries of $\bm o$ and $\bm \theta$ are real and positive.
In addition, according to the Schur-Horn theorem, $\bm o$ is majorized by $\bm \theta$, i.e., 
\begin{equation}\label{majorization}
\bm o \prec \bm \theta.
\end{equation}
Define a real vector $\bm u = ( u_1, \cdots, u_K )^T$ with $u_k > 0, ~\forall~ k \in {\cal K}$, and function $g_2 (\bm u) = \sum_{k=1}^K \frac{1}{u_k}$.
It is obvious that $g_2 (\bm u)$ is convex and symmetric.
Hence, $g_2 (\bm u)$ is a Schur-convex function.
Therefore, 
\begin{equation}\label{ga_glambda}
g_2 (\bm o) \leq g_2 (\bm \theta).
\end{equation}
Using (\ref{ga_glambda}), we have
\begin{align}
{\text {tr}} \left( {\bm O}_1^{-1} \right) & = \sum_{k=1}^K \frac{1}{o_k}\nonumber\\
&= g_2 (\bm o) \nonumber\\
& \leq g_2 (\bm \theta)\nonumber\\
& = \sum_{k=1}^K \frac{1}{\theta_k}\nonumber\\
& = {\text {tr}} \left( \bm O^{-1} \right),
\end{align}
based on which we get $g_1'(x) \leq 0$ and (\ref{ineq1}) can then be proven.
\hfill $\Box$

Then, from (\ref{R_lb1_C_infty}), (\ref{Capacity_channel_1}), and Lemma \ref{mutuan_ine}, it is known that when $C \rightarrow + \infty$,
\begin{align}
R^{\text {lb}1} &  \rightarrow  \mathbb{E} \left[ \log \det \left( \bm I_K + \bm A_1 \right) - \log \det \left( \bm A_1 \right) \right] \nonumber\\
& =  K \mathbb{E} \left[ \log \left( 1 + \frac{1}{a} \right) \right] \nonumber\\
& \leq I(\bm x; \bm y, \bm H),
\end{align}
where the expectation can be calculated by using the pdf of $a$ in (\ref{pdf_a}).
Lemma \ref{lemma_QCI} is thus proven.

\section{Proof of Theorem \ref{theorem_MMSE}}
\label{prove_theorem_MMSE}

As stated in Appendix \ref{prove_theorem_ub}, $\bm U \bm \varLambda \bm U^H$ is the eigendecomposition of $\bm H \bm H^H$ and $\lambda_t, \forall t \in {\cal T}$ are unordered positive eigenvalues of $\bm H \bm H^H$.
To derive $R^{\text {lb}2}$, we further denote the singular value decomposition of $\bm H$ by $\bm U \bm L \bm V^H$, where 
$\bm V \in {\mathbb C}^{K \times K}$ is a unitary matrix and $\bm L \in {\mathbb R}^{M \times K}$ is a rectangular diagonal matrix.
In fact, the diagonal entries of $\bm L$ are the non-negative square roots of the positive eigenvalues of $ \bm H \bm H^H$.
Then, from (\ref{F}), we have
\begin{align}\label{FH_FF}
& \bm F^H \bm H \nonumber\\
= & \bm H^H \left( \bm H \bm H^H + \sigma^2 \bm I_M \right)^{-1} \bm H, \nonumber\\
= & \bm V \bm L^H \left( \bm \varLambda + \sigma^2 \bm I_M \right)^{-1} \bm L \bm V^H, \nonumber\\
= & \bm V {\text {diag}} \left\{ \frac{\lambda_1}{\lambda_1 + \sigma^2}, \cdots, \frac{\lambda_T}{\lambda_T + \sigma^2}, \bm 0_{K-T}^H \right\} \bm V^H, \nonumber\\
& \bm F^H \bm H \bm H^H \bm F \nonumber\\
= & \bm V \bm L^H \left( \bm \varLambda + \sigma^2 \bm I_M \right)^{-1} \bm \varLambda \left( \bm \varLambda + \sigma^2 \bm I_M \right)^{-1} \bm L \bm V^H, \nonumber\\
= & \bm V {\text {diag}} \left\{ \frac{\lambda_1^2}{\left(\lambda_1 + \sigma^2\right)^2}, \cdots, \frac{\lambda_T^2}{\left(\lambda_T + \sigma^2\right)^2}, \bm 0_{K-T}^H \right\} \bm V^H, \nonumber\\
& \bm F^H \bm F \nonumber\\
= & \bm V \bm L^H \left( \bm \varLambda + \sigma^2 \bm I_M \right)^{-2} \bm L \bm V^H,\nonumber\\
= & \bm V {\text {diag}} \left\{ \frac{\lambda_1}{\left(\lambda_1 + \sigma^2\right)^2}, \cdots, \frac{\lambda_T}{\left(\lambda_T + \sigma^2\right)^2}, \bm 0_{K-T}^H \right\} \bm V^H,
\end{align}
where $\bm 0_{K-T}$ is a $(K-T)$-dimensional all `$0$' column vector.
Based on (\ref{FH_FF}), 
\begin{align}\label{FHHF_FF}
& \bm F^H \bm H \bm H^H \bm F + \sigma^2 \bm F^H \bm F + D \bm I_K \nonumber\\
= & \bm V {\text {diag}} \left\{ \frac{\lambda_1}{\lambda_1 + \sigma^2} + D, \cdots, \frac{\lambda_T}{\lambda_T + \sigma^2} + D, D \times \bm 1_{K-T}^H \right\} \bm V^H,
\end{align}
where $\bm 1_{K-T}$ is a $(K-T)$-dimensional all `$1$' column vector.
Since $\bm \varLambda$ is independent of $\bm U$, $\bm L$ is independent of $\bm U$ as well as $\bm V$, and $\lambda_t, \forall t \in {\cal T}$ are unordered, we have
\begin{align}\label{exp_FHHF_FF}
& {\mathbb E} \left[ \log \det \left(\bm F^H \bm H \bm H^H \bm F + \sigma^2 \bm F^H \bm F + D \bm I_K\right) \right] \nonumber\\
= & T {\mathbb E} \left[ \log \left( \frac{\lambda}{\lambda + \sigma^2} + D\right) \right] + (K-T) \log D.
\end{align}
Then, we calculate $\bm G$ in (\ref{G}).
For this purpose, we have to calculate ${\mathbb E} \left[ \bm F^H \bm H \right]$, ${\mathbb E} \left[ \bm F^H \bm H \bm H^H \bm F \right]$, and ${\mathbb E} \left[ \bm F^H \bm F \right]$.
To get these expectations, we consider two different cases, i.e., the case with $K \leq M$ and the case with $K>M$.
When $K \leq M$, from (\ref{FH_FF}), we have
\begin{align}\label{exp1}
& {\mathbb E} \left[ \bm F^H \bm H \right] = {\mathbb E} \left[ \frac{\lambda}{\lambda + \sigma^2} \right] \bm I_K,\nonumber\\
& {\mathbb E} \left[ \bm F^H \bm H \bm H^H \bm F \right] = {\mathbb E} \left[ \frac{\lambda^2}{(\lambda + \sigma^2)^2} \right] \bm I_K,\nonumber\\
& {\mathbb E} \left[ \bm F^H \bm F \right] = {\mathbb E} \left[ \frac{\lambda}{(\lambda + \sigma^2)^2} \right] \bm I_K.
\end{align}
When $K>M$, denote $\bm V = (\bm v_1, \cdots, \bm v_K)$.
Then, from (\ref{FH_FF}),
\begin{align}\label{FH}
\bm F^H \bm H & = \bm V {\text {diag}} \left\{ \frac{\lambda_1}{\lambda_1 + \sigma^2}, \cdots, \frac{\lambda_M}{\lambda_M + \sigma^2}, \bm 0_{K-T}^H \right\} \bm V^H\nonumber\\
& = \left( \frac{\lambda_1}{\lambda_1 + \sigma^2} \bm v_1, \cdots, \frac{\lambda_M}{\lambda_M + \sigma^2} \bm v_M, \bm 0_K^H, \cdots, \bm 0_K^H \right) \begin{bmatrix}
\bm v_1^H\\
\vdots   \\
\bm v_K^H\\
\end{bmatrix}\nonumber\\
& = \sum_{m=1}^M \frac{\lambda_m}{\lambda_m + \sigma^2} \bm v_m \bm v_m^H.
\end{align}
Since $\bm v_m$ is the eigenvector of matrix $\bm H^H \bm H$ and is independent of unordered eigenvalue $\lambda_m$, we have
\begin{align}\label{exp_FH}
{\mathbb E} \left[ \bm F^H \bm H \right] & = \sum_{m=1}^M {\mathbb E} \left[ \frac{\lambda_m}{\lambda_m + \sigma^2} \right] \frac{1}{K} \bm I_K \nonumber\\
& = \frac{M}{K} {\mathbb E} \left[ \frac{\lambda}{\lambda + \sigma^2} \right] \bm I_K.
\end{align}
Similarly, we also have 
\begin{align}\label{exp_FHHF}
& {\mathbb E} \left[ \bm F^H \bm H \bm H^H \bm F \right] = \frac{M}{K} {\mathbb E} \left[ \frac{\lambda^2}{(\lambda + \sigma^2)^2} \right] \bm I_K, \nonumber\\
& {\mathbb E} \left[ \bm F^H \bm F \right] = \frac{M}{K} {\mathbb E} \left[ \frac{\lambda}{(\lambda + \sigma^2)^2} \right] \bm I_K.
\end{align}
Using (\ref{exp1}), (\ref{exp_FH}), (\ref{exp_FHHF}), and (\ref{G}), $\bm G$ can be calculated as
\begin{align}\label{G2}
\bm G & = {\mathbb E} \left[ \bm F^H \bm H \bm H^H \bm F \right] - {\mathbb E} \left[ \bm F^H \bm H \right]  {\mathbb E} \left[ \bm H^H \bm F \right]\nonumber\\
& + \sigma^2 {\mathbb E} \left[ \bm F^H \bm F \right] + D \bm I_K\nonumber\\
& = \left\{ \frac{T}{K} {\mathbb E} \left[ \frac{\lambda}{\lambda + \sigma^2} \right] - \frac{T^2}{K^2} \left( {\mathbb E} \left[ \frac{\lambda}{\lambda + \sigma^2} \right] \right)^2 + D \right\} \bm I_K.
\end{align}
Hence, 
\begin{align}\label{log_det_G}
& \log \det (\bm G)  \nonumber\\
= & K \log \left\{ \frac{T}{K} {\mathbb E} \left[ \frac{\lambda}{\lambda + \sigma^2} \right] - \frac{T^2}{K^2} \left( {\mathbb E} \left[ \frac{\lambda}{\lambda + \sigma^2} \right] \right)^2 + D \right\}.
\end{align}
Substituting (\ref{exp_FHHF_FF}) and (\ref{log_det_G}) into (\ref{diff_z_H_lb1}) and (\ref{diff_z_x_lb1}), respectively, and using (\ref{mutual_x_z2}), we can get (\ref{R_lb4}).

We then calculate $D$ in (\ref{D_2}).
From (\ref{bar_y_lb4}), (\ref{exp1}), and (\ref{exp_FHHF}),
\begin{align}
{\mathbb E} \left[ {\bar {\bm x}} {\bar {\bm x}}^H \right] & = {\mathbb E} \left[ \bm F^H \bm H \bm H^H \bm F + \sigma^2  \bm F^H \bm F \right]\nonumber\\
& = \frac{T}{K} {\mathbb E} \left[ \frac{\lambda}{\lambda + \sigma^2} \right] \bm I_K.
\end{align}
$I({\bar {\bm x}}_g; {\bar {\bm z}}_g)$ in (\ref{mutual_yg_zg_lb1}) can thus be calculated as follows
\begin{align}\label{mutual_yg_zg_lb1_5}
I({\bar {\bm x}}_g; {\bar {\bm z}}_g) & = \log \det \left( \bm I_K + \frac{{\mathbb E} \left[ {\bar {\bm x}} {\bar {\bm x}}^H \right]}{D} \right) \nonumber\\
& = K \log \left( 1 + \frac{T}{DK} {\mathbb E} \left[ \frac{\lambda}{\lambda + \sigma^2} \right] \right) \nonumber\\
& = C,
\end{align}
based on which (\ref{D_2}) can be obtained.
Theorem \ref{theorem_MMSE} is then proven.

\section{Proof of Lemma \ref{lemma_MMSE}}
\label{prove_lemma_MMSE}

When $M \rightarrow + \infty$, $T=K$. 
As stated in Appendix \ref{prove_lemma_ub}, $\bm H^H \bm H - M \bm I_K \rightarrow \bm 0$ almost surely.
Hence, $\lambda - M \rightarrow 0$.
From (\ref{mutual_yg_zg_lb1_5}),
\begin{align}\label{mutual_yg_zg_lb1_2}
I({\bar {\bm x}}_g; {\bar {\bm z}}_g) & = K \log \left( 1 + \frac{1}{D} {\mathbb E} \left[ \frac{\lambda}{\lambda + \sigma^2} \right] \right) \nonumber\\
& = C \nonumber\\
& \rightarrow K \log \left( 1+\frac{1}{D} \right).
\end{align}
Combining (\ref{R_lb4}) and (\ref{mutual_yg_zg_lb1_2}), we have
\begin{align}\label{R_lb4_infi}
R^{\text {lb}2} & \rightarrow K \log (1+D)- K \log D\nonumber\\
& = K \log \left( 1 + \frac{1}{D} \right) \nonumber\\
& \rightarrow C.
\end{align}

When $K \leq M$ and $\rho \rightarrow + \infty$, $T=K$ and $\sigma^2 \rightarrow 0$.
Using (\ref{mutual_yg_zg_lb1_5}) and (\ref{R_lb4}), we can also get (\ref{mutual_yg_zg_lb1_2}) and (\ref{R_lb4_infi}).

When $K \leq M$ and $C \rightarrow + \infty$, it can be found from (\ref{D_2}) that $D \rightarrow 0$.
Then, using (\ref{R_lb4}), we can get (\ref{R_lb4_appro}).
This finishes the proof.

\section*{Acknowledgments}
This work was supported by the Alexander von Humboldt Foundation and the work of S. Shamai has been supported by the
European Union's Horizon 2020 Research and Innovation Programme with grant agreement No. 694630.

\bibliographystyle{IEEEtran}
\bibliography{IEEEabrv,Ref}

\begin{thebibliography}{10}
\providecommand{\url}[1]{#1}
\csname url@samestyle\endcsname
\providecommand{\newblock}{\relax}
\providecommand{\bibinfo}[2]{#2}
\providecommand{\BIBentrySTDinterwordspacing}{\spaceskip=0pt\relax}
\providecommand{\BIBentryALTinterwordstretchfactor}{4}
\providecommand{\BIBentryALTinterwordspacing}{\spaceskip=\fontdimen2\font plus
\BIBentryALTinterwordstretchfactor\fontdimen3\font minus
  \fontdimen4\font\relax}
\providecommand{\BIBforeignlanguage}[2]{{%
\expandafter\ifx\csname l@#1\endcsname\relax
\typeout{** WARNING: IEEEtran.bst: No hyphenation pattern has been}%
\typeout{** loaded for the language `#1'. Using the pattern for}%
\typeout{** the default language instead.}%
\else
\language=\csname l@#1\endcsname
\fi
#2}}
\providecommand{\BIBdecl}{\relax}
\BIBdecl

\bibitem{tishby2000information}
N.~Tishby, F.~C. Pereira, and W.~Bialek, ``The information bottleneck method,''
  \emph{arXiv preprint physics/0004057}, 2000.

\bibitem{shwartz2017opening}
R.~Shwartz-Ziv and N.~Tishby, ``Opening the black box of deep neural networks
  via information,'' \emph{arXiv preprint arXiv:1703.00810}, 2017.

\bibitem{dobrushin1962information}
R.~Dobrushin and B.~Tsybakov, ``Information transmission with additional
  noise,'' \emph{IRE Trans. Inf. Theory}, vol.~8, no.~5, pp. 293--304, Sep.
  1962.

\bibitem{witsenhausen1980indirect}
H.~Witsenhausen, ``Indirect rate distortion problems,'' \emph{IEEE Trans. Inf.
  Theory}, vol.~26, no.~5, pp. 518--521, Sep. 1980.

\bibitem{courtade2013multiterminal}
T.~A. Courtade and T.~Weissman, ``Multiterminal source coding under logarithmic
  loss,'' \emph{IEEE Trans. Inf. Theory}, vol.~60, no.~1, pp. 740--761, Jan.
  2014.

\bibitem{aguerri2019capacity}
I.~E. Aguerri, A.~Zaidi, G.~Caire, and S.~S. Shitz, ``On the capacity of cloud
  radio access networks with oblivious relaying,'' \emph{IEEE Trans. Inf.
  Theory}, vol.~65, no.~7, pp. 4575--4596, July 2019.

\bibitem{nazer2011compute}
B.~Nazer and M.~Gastpar, ``Compute-and-forward: Harnessing interference through
  structured codes,'' \emph{IEEE Trans. Inf. Theory}, vol.~57, no.~10, pp.
  6463--6486, Oct. 2011.

\bibitem{hong2013compute}
S.-N. Hong and G.~Caire, ``Compute-and-forward strategies for cooperative
  distributed antenna systems,'' \emph{IEEE Trans. Inf. Theory}, vol.~59,
  no.~9, pp. 5227--5243, Sep. 2013.

\bibitem{nazer2009structured}
B.~Nazer, A.~Sanderovich, M.~Gastpar, and S.~Shamai, ``Structured superposition
  for backhaul constrained cellular uplink,'' in \emph{Proc. IEEE Int. Symp.
  Inf. Theory (ISIT)}, Seoul, South Korea, June 2009, pp. 1530--1534.

\bibitem{simeone2011codebook}
O.~Simeone, E.~Erkip, and S.~Shamai, ``On codebook information for interference
  relay channels with out-of-band relaying,'' \emph{IEEE Trans. Inf. Theory},
  vol.~57, no.~5, pp. 2880--2888, May 2011.

\bibitem{park2012robust}
S.-H. Park, O.~Simeone, O.~Sahin, and S.~Shamai, ``Robust and efficient
  distributed compression for cloud radio access networks,'' \emph{IEEE Trans.
  Veh. Technol.}, vol.~62, no.~2, pp. 692--703, Feb. 2013.

\bibitem{zhou2016optimal}
Y.~Zhou, Y.~Xu, W.~Yu, and J.~Chen, ``On the optimal fronthaul compression and
  decoding strategies for uplink cloud radio access networks,'' \emph{IEEE
  Trans. Inf. Theory}, vol.~62, no.~12, pp. 7402--7418, Dec. 2016.

\bibitem{aguerri2016lossy}
I.~E. Aguerri and A.~Zaidi, ``Lossy compression for compute-and-forward in
  limited backhaul uplink multicell processing,'' \emph{IEEE Trans. Commun.},
  vol.~64, no.~12, pp. 5227--5238, Dec. 2016.

\bibitem{demel2020cloud}
J.~Demel, T.~Monsees, C.~Bockelmann, D.~Wuebben, and A.~Dekorsy, ``Cloud-ran
  fronthaul rate reduction via ibm-based quantization for multicarrier
  systems,'' in \emph{Proc. 24th International ITG Workshop on Smart Antennas},
  Hamburg, Germany, Feb. 2020, pp. 1--6.

\bibitem{winkelbauer2014rate}
A.~Winkelbauer and G.~Matz, ``Rate-information-optimal {Gaussian} channel
  output compression,'' in \emph{Proc. 48th Annu. Conf. Inf. Sci. Syst.
  (CISS)}, Princeton, NJ, USA, Mar. 2014, pp. 1--5.

\bibitem{winkelbauer2014ratevec}
A.~Winkelbauer, S.~Farthofer, and G.~Matz, ``The rate-information trade-off for
  {Gaussian} vector channels,'' in \emph{Proc. IEEE Int. Symp. Inf. Theory},
  Honolulu, USA, June 2014, pp. 2849--2853.

\bibitem{caire2018information}
G.~Caire, S.~Shamai, A.~Tulino, S.~Verdu, and C.~Yapar, ``Information
  bottleneck for an oblivious relay with channel state information: the scalar
  case,'' in \emph{Proc. IEEE Int. Conf. Science of Electrical Engineering in
  Israel (ICSEE)}, Eilat, Israel, Dec. 2018, pp. 1--5.

\bibitem{cover2012elements}
T.~M. Cover and J.~A. Thomas, \emph{Elements of information theory}.\hskip 1em
  plus 0.5em minus 0.4em\relax John Wiley \& Sons, 2012.

\bibitem{telatar1999capacity}
E.~Telatar, ``Capacity of multi-antenna gaussian channels,'' \emph{Europ.
  Trans. Telecommun.}, vol.~10, no.~6, pp. 585--595, Nov.-Dec. 1999.

\bibitem{tulino2004random}
A.~M. Tulino, S.~Verd{\'u} \emph{et~al.}, \emph{Random matrix theory and
  wireless communications}.\hskip 1em plus 0.5em minus 0.4em\relax {Now
  Publishers}, 2004.

\bibitem{lee1990estimate}
W.~C. Lee, ``Estimate of channel capacity in rayleigh fading environment,''
  \emph{IEEE trans. Veh. Tech.}, vol.~39, no.~3, pp. 187--189, Aug. 1990.

\bibitem{brennan1982adaptive}
L.~E. Brennan and I.~S. Reed, ``An adaptive array signal processing algorithm
  for communications,'' \emph{IEEE Trans. Aerosp. Electron. Syst.}, no.~1, pp.
  124--130, Jan. 1982.

\bibitem{csiszar1992arbitrarily}
I.~Csiszar, ``Arbitrarily varying channels with general alphabets and states,''
  \emph{IEEE Trans. Inf. Theory}, vol.~38, no.~6, pp. 1725--1742, Nov. 1992.

\end{thebibliography}
\end{document}